\documentclass[aps,11pt,tightenlines,onecolumn,showpacs,pra,citeautoscript,amsmath,amssymb,floatfix,superscriptaddress,nofootinbib]{revtex4-2}

\usepackage{amsmath}
\usepackage{amssymb}
\usepackage{epsfig}
\usepackage{color,epstopdf}
\usepackage{amsthm, thm-restate}
\usepackage{hyperref}
\usepackage{csquotes}
\usepackage{mathtools}
\usepackage{xcolor}
\usepackage{comment}
\usepackage{float}
	
\usepackage[normalem]{ulem}
\usepackage[caption=false]{subfig}
\usepackage{algpseudocode}

\usepackage{tcolorbox}

\usepackage{paralist}

\newtheorem*{theorem*}{Theorem}

\newtheorem*{lemma*}{Lemma}

\makeatletter
\renewcommand*\l@section{\@dottedtocline{1}{1.5em}{2em}}
\renewcommand*\l@subsection{\@dottedtocline{1}{4em}{2em}}
\makeatother

\begin{document}
\count\footins = 1000
\title{Algorithmic idealism: what should you believe to experience next?}

\author{Markus P.\ M\"uller}
\email{Markus.Mueller@oeaw.ac.at}
\affiliation{Institute for Quantum Optics and Quantum Information,
Austrian Academy of Sciences, Boltzmanngasse 3, A-1090 Vienna, Austria}
\affiliation{Vienna Center for Quantum Science and Technology (VCQ), Faculty of Physics, University of Vienna, Vienna, Austria}
\affiliation{Perimeter Institute for Theoretical Physics, 31 Caroline Street North, Waterloo, Ontario N2L 2Y5, Canada}

\date{July 24, 2025}

\begin{abstract}
I argue for an approach to the Foundations of Physics that puts the question in the title center stage, rather than asking ``what is the case in the world?''. This approach, algorithmic idealism, attempts to give a mathematically rigorous \textit{in-principle}-answer to this question both in the usual empirical regime of physics and in some more exotic regimes within cosmology, philosophy, and science fiction (but soon perhaps real) technology. I begin by arguing that quantum theory, in its actual practice and in some interpretations, should be understood as telling an agent what they should expect to observe next (rather than what is the case), and that the difficulty of answering this former question from the usual ``external'' perspective is at the heart of persistent enigmas such as the Boltzmann brain problem, extended Wigner's friend scenarios, Parfit's teletransportation paradox, or our understanding of the simulation hypothesis. Algorithmic idealism is a conceptual framework, based on two postulates that admit several possible mathematical formalizations, cast in the language of algorithmic information theory. Here I give a non-technical description of this view and show how it dissolves the aforementioned enigmas: for example, it claims that you should never bet on being a Boltzmann brain regardless of how many there are, that shutting down computer simulations does not generally terminate its inhabitants, and it predicts the apparent embedding into an objective external world as an approximate description.
\end{abstract}

\maketitle

\tableofcontents

\section{Prologue: when the world is not enough}
\label{SecPrologue}
It is the year 2048. You can no longer feel your arms and legs, but you dream of the warm autumn sun breaking through colorful leaves. A blackbird calls from a distance and your daughter smiles at you, a picnic blanket, the shade of the trees. Only the flickering of the ceiling light and the flashing of the surveillance monitors bring you back to reality, and only temporarily, until the warm feeling of the pain medication makes your perception fade.

You are terminally ill. You only have a few days to live, at least that's what the Doctor says. And yet this realization, in the bright moments between the effects of the morphine and drifting off to sleep, does not fill you with despair: you have taken precautions. Two years ago you signed a contract with AfterMath Ltd.: shortly after your death you will be scanned\footnote{This story and most of what follows assumes that agents (at least the ones we mean here, which is ultimately supposed to include humans) can in principle be described classically, and potentially with a finite amount of classical information. This is compatible with the fact that some biological processes are genuinely quantum-mechanical and / or continuous, but a possible intuition is that these details must ultimately be irrelevant for all that matters for us as persons (including conscious experience), since robust functioning seems to forbid dependence on too fine-grained details. Whether this assumption is well-justified is the subject of contemporary debate, with a vast body of literature on e.g.\ the (un)suitability of ``mind uploading''. For a recent proposal against this assumption, claiming that quantum states are at the heart of consciousness, see e.g.~\cite{ArianoFaggin}.} by a team of specially trained neuroscientists with a particular machine, and your body will be destroyed in the process. A few days later, you will be digitally resurrected in a simulated world on a computer.

You have not made this decision lightly. You have spent years studying philosophy, ethics and neuroscience. In the end, it was thoughts like Bostrom's fable~\cite{Bostroem2005} and the experimental results of AfterMath that convinced you to go for it. You are pretty sure that you are trying the right thing. And you are hopeful that it will go well.

And yet, you are afraid. You are afraid because you are human. The perception fades, the pain disappears, but the fear and the will to live remain. Your eyes are closed, but you feel the Doctor enter the room. You must have heard his feet scrubbing across the plastic floor of the hospital room, albeit unconsciously.

You clear your throat. At first you can't get a word out, but then you manage to whisper quietly.

\textbf{You:} \textit{``Doctor, I'm so scared...''} (coughing) \textit{``I know this is not your specialty, but... I need to know! Will I really wake up in the computer simulation?''}

You instantly regret having asked, because you know the doctor very well. The doctor is a former physicist, not only a physician, but a \textit{physicalist} by conviction. He is excellent in his job, but you don't remember him as particularly empathetic.

\textbf{Doctor:} \textit{``Hahaha, you fool! You are asking a non-question! All there is to say is that there is a human being here now, and a computer running a simulation of that thing later. This is all there is to know about the facts of the world.''}

The Doctor goes on to refill your infusion. \textit{``You've paid AfterMath to run that computer simulation, and this is what is going to happen. I don't even understand what you are actually uncertain \underline{about}? You know exactly what will be the case in the world!}''\\

\begin{tcolorbox}[width=\linewidth, sharp corners=all, colback=white!95!black]
Please note that this is a \textbf{summary and motivation} of algorithmic idealism, intentionally dropping many details. For the precise mathematical derivations and conceptual definitions, please see the publications~\cite{Mueller2020,JonesMueller}.
\end{tcolorbox}

\section{Quantum probabilities as objective degrees of epistemic justification}
\label{SecQuantum}

We usually think that physics is the science of what the world is like. The idea that the contents of our physical theories represent actual things in the world, and that these things have properties that exist objectively and independently of any observer, is a characteristic assumption of many versions of scientific realism~\cite{SE-ScientificRealism}. And there are good reasons to be a scientific realist: for example, realism allows us to understand that the success of science is not a miracle, but due to the fact that our theories are at least approximately representative of the truth. Intuitively, rejecting realism seems dangerously close to irrationality and pseudoscience~\cite{SokalBricmont}. And yet, quantum physics challenges\footnote{To avoid potential misunderstandings from the outset, \textbf{algorithmic idealism is not meant to be a new interpretation of quantum theory}. It is a theory that aims to make concrete predictions in regimes where they are currently lacking (``private experiments'' as discussed in Section~\ref{SecRestrictionA}). In order to do so, it turns out necessary to set standard realist assumptions aside and focus exclusively on the question of the title of this paper. Quantum physics is seen as additional motivation for this focus. The goal is to construct the theory and explore its predictions (some of which are highly counterintuitive), not to have a detailed and philosophically exhaustive discussion of quantum theory, probability, or of all conceptual views that are articulated as motivation for it.} at least the most naive versions of realism.

This is evident in the scientific practice of quantum physics. Imagine a particle that has been prepared in a superposition state of two paths in an interferometer. When we measure, we will find the particle in one of the branches, and not in the other. But what is the case before the measurement? A naive stipulation that the particle is in one of the branches, but we do not know which one, leads easily to contradictions with the actual statistics we observe in experiments\footnote{In more detail, a contradiction appears if we have a final beamsplitter that leads to perfect destructive interference, \textit{and} if we naively model the experiment as a probability-$1/2$ mixture of the two experimental realizations where the particle is deterministically prepared in the upper or in the lower branch, respectively. It is simply a methodological observation that the most straightforward epistemic interpretations tend to fail to explain the experimental results. Making this simple methodological observation formally rigorous is important, but requires substantially more work, e.g.\ of proving that no noncontextual ontological model reproduces the specific functional form of the quantum uncertainty relations characteristic of the interference experiment~\cite{Catani2022}.}.

We can certainly write down a wavefunction that describes all there is to say about the particle before the measurement, but as soon as we claim that the wavefunction represents ``what is actually the case'', there appears a strange disconnect between the facts that we actually see (the concrete ``classical'' measurement outcomes) and those before we look (the quantum states). Wavefunctions do not satisfy our basic intuitions about how ``facts of the world'' should behave. For example, it is provably impossible to observe them directly~\cite{WoottersZurek}. If we think of the wave functions assigned by some observer as a real property of the physical system that it is supposed to describe, then it would seem to collapse nonlocally and instantaneously upon measurement, in potential tension with relativity (see e.g.~\cite[Section 6.1]{Peres} for more details)\footnote{These observations certainly do not rule out the possibility of realist interpretations of the quantum state. For example, Everettian interpretations evade the need for collapse, at the expense of distinguishing the \textit{absolute} quantum state of the universe from the branch-\textit{relative} quantum state assigned by an observer~\cite{Everett}. This is arguably a significant departure from our intuitive view on ``facts of the world''. De Broglie-Bohm theory and objective collapse models are further examples of consistent realist interpretations, but whether they can be (or have been already) formulated in a way that avoids conflict with special relativity is subject to ongoing debate~\cite{Duerr,SEPCollapse}.}.

Methodologically, this implies that quantum physics does not allow us to rely on our intuitions shaped by naive realism about how things in the world and their properties should behave. Now, the quantum foundations community has rightly pointed out that a lack of intuition, or of imagination, is insufficient to draw any direct metaphysical conclusions. Many phenomena that seem genuinely quantum, such as interference, entanglement, no-cloning, or teleportation, can be reproduced ``classically'', i.e.\ by models based on classical variables that satisfy natural properties such as locality or noncontextuality~\cite{Spekkens2007,Hausmann2021,Catani2023}. What is needed are no-go theorems that tell us that it is \textit{provably impossible} to reproduce a phenomenon classically under such assumptions --- but those results exist, and Bell's theorem~\cite{Bell} is the most paradigmatic example of this. It implies that the correlations between spacelike separated events cannot always be explained by a classical common cause, i.e.\ by a classical hidden-variable model that respects the causal structure of the experimental setup~\cite{WoodSpekkens}. Hence, our realist intuition that the local measurements simply uncover objective, preexisting facts of the world must go wrong, unless locality is violated. In scientific practice, it is then often a methodological move to give up on realist intuitions for quantum systems ``in between'' measurements.

Acknowledging the limitations of the received realist view becomes actually \textit{explanatorily powerful}: it allows us, for example, to understand intuitively why ``device-independent cryptography'' works~\cite{BarrettHardyKent,Colbeck,Pironio}. In this scheme, Alice and Bob perform measurements on two quantum systems that have been prepared in an entangled state, and observe a violation of a Bell inequality on many repetitions. Without any knowledge of the workings of their devices, this allows them in some cases to extract private random bits that are provably unpredictable by any adversary subject to the no-signalling principle. Intuitively, this is because the random bits have not been ``facts of the world'' before the measurements, and nobody can spy on non-existent bits.

The difficulty of saying what quantum states tell us about the world invites us hence to shift the focus from the standard realist view towards a more empiricist perspective. We can focus on what quantum states \textit{definitely do} tell us, according to the consensus of the practicing physicists: they tell us \textit{what we should expect to observe in an experiment}. Given a specification of an experiment which implies a description of the involved quantum states and measurement operators, the Born rule tells us which probabilities to assign to the possible measurement outcomes. In a double-slit experiment, for example, it would be wrong to think that the quantum state tells us through which slit the particle is actually passing; but it tells us what to expect about the place at which we will observe the particle hitting the screen. In other words, Quantum Theory is about \textit{what we should believe to observe next}.

This methodological diagnosis is compatible with many interpretations of Quantum Theory\footnote{This diagnosis is also compatible with QBism, even if the QBists would perhaps not like to phrase it in these terms. They would emphasize that the Born rule is a mere \textit{consistency} requirement~\cite{DeBrota} between an agent's personal assignments of the quantum state, the measurement operator, and the outcome probability. But this implies that any agent that has made up its mind about what quantum state and measurement operator to assign \textit{should} hold beliefs about outcome probabilities that are given by the Born rule.}, since they all agree on how to practically \textit{use} Quantum Theory successfully. But we can go one step further, and claim that \textit{this is actually what the quantum state is}. This is the content of Berghofer's \textit{degrees of epistemic justification interpretation} (DEJI)~\cite{Berghofer}: \textit{``In short, DEJI is an agent-centered interpretation that views quantum mechanics as a single-user theory that allows an experiencing subject to answer the following question: Based on my experiential input, what should I believe to experience next? The input is the wave function and the output is quantum probabilities.''}

Berghofer's interpretation\footnote{Berghofer's interpretation and algorithmic idealism differ significantly: DEJI is an interpretation of quantum theory, and algorithmic idealism (AI) is not. Quantum theory matters for AI in two ways: first, as a motivation to focus on the question of what is observed next; and second, as a ``litmus test'' to see whether some quantum phenomena are predicted by AI. However, AI is not an interpretation of an existing theory, but a new theory that intends to make novel predictions, as outlined in the next sections. AI benefits from the conceptually rigorous formulation of DEJI, which clarifies its own use of probability substantially.} can be understood as a modification of QBism~\cite{Fuchs2017} (formerly known as Quantum Bayesianism), which views quantum states as subjective degrees of belief. However, given that quantum mechanics is our most successful and fundamental theory of physics, this raises the question of ``how objectivity could enter science'' in QBism~\cite{Berghofer}. According to DEJI, quantum probabilities are objective degrees of epistemic justification. In this sense, quantum probabilities tell us something objective about the world: not what is the case in the world, but what we \textit{should} expect Nature to answer if we ask it a question. In the rest of this paper, we will consider several exotic situations in which agents ask Nature a question. We are not interested in what agents \textit{believe}, but what type of laws determines what they will \textit{actually probably} receive as an answer --- or, in other words, what they \textit{should} believe to observe. Hence, Berghofer's work will be our guiding interpretation when formulating such probabilistic laws.

\section{Restriction A: physics does not always tell agents what they should believe}
\label{SecRestrictionA}

Berghofer's interpretation is but one example of a broader class of ``neo-Bohrian''~\cite{Cuffaro} interpretations, which include Healey's pragmatism~\cite{Healey}, Brukner and Zeilinger's~\cite{BruknerZeilinger} views, Bub's information-theoretic interpretation~\cite{Bub}, Rovelli's relational interpretation~\cite{Rovelli}, and others. The exact differences between these views, and how to best delineate them from other views and each other, are not relevant for the purpose of this paper. The conclusion that I argue to be drawn at this point is simply this: \textit{both scientific practice and a class of interpretations (most explicitly, Berghofer's) suggest that Quantum Theory is best understood as telling an agent what they should believe to experience next}.

It is not necessary to agree with any one of these interpretations to share the broadly empiricist sentiment that the question of ``What will I see next?'' is an epistemically more modest and perhaps more fruitful one than ``What is really the case in the world?''. After all, the world might be a much weirder place than we have ever imagined, and humans have a bad track record of guessing the counterintuitive metaphysics of modern physics a priori. Methodologically, it makes sense to concentrate on the requirement that physics should allow us, at the very least, to make successful predictions. This includes predictions that can be intersubjectively verified, but also \textit{private} ones that are in some sense irreducibly relative to an observer\footnote{Physical theories do not directly predict \textit{all} intersubjectively communicable outcomes of experiments, but only those that are in some precise sense objectively definable; and similarly, algorithmic idealism does not aim for predictions of \textit{all} outcomes of private experiments. For example, physics allows us to predict probabilistically whether we will hear a detector click, and algorithmic idealism says that we should be able to do the same even if the outcome is private (as in a duplication or Wigner's friend type experiment). However, neither physics nor algorithmic idealism address directly questions of qualia, say, whether e.g.\ an agent will observe beauty or joy or be conscious next. We may certainly believe that these \textit{supervene} on the precisely definable properties, but to understand \textit{how exactly} this is the case might be so difficult to characterize that it may be a better strategy to consider the study of these question a different field of inquiry.}.

It may at first sight seem unlikely that there should be any relevant sought-for predictions of this ``private'' type, but here I will argue that there are in fact plenty, and that some of them are of utmost importance for physics, philosophy, and humanity at large. The question raised by the patient in Section~\ref{SecPrologue} is an archetypical example of this. The answer to the patient's question \textit{``Will I really wake up in the computer simulation?''} can only be verified privately, not intersubjectively. That is, any claim of an answer cannot be empirically tested \textit{externally}; say, by running the scanning and simulation process on a large number of people, and doing statistics about how many times the respective test person has \textit{actually} woken up in the simulation. The answer to the question is by construction \textit{independent} of any facts of the world\footnote{The problem here is much deeper than the use of indexicals per se. For example, the question of \textit{``Where am I?''} can be rephrased as asking \textit{``Where is this particular patient [insert name and ID number, or some other unique external identifier] currently located?''}, i.e.\ deindexicalized, and an answer can be derived from facts of the world. In Adlam's~\cite{Adlam} terminology, we can always use our beliefs about what is the case in the world to set our \textit{superficially self-locating credences}, in contrast to \textit{pure self-locating credences} which are subject to Restriction A.} that any third-person view could discover\footnote{We can certainly imagine a (contrived) theory that claims, for example, that the patient will experience waking up in the simulation if and only if the temperature in the hospital room was less than 30$^{\circ}$C at the time of scanning. Then learning a third-person fact (the temperature) \textit{would} tell us the answer to this first-person question, \textit{assuming that this theory is true}. However, simply learning the temperature and inferring all consequences of this \textit{as predicted by our current physical theories}, or as predicted by any other future physical theory that makes only intersubjectively verifiable claims, does not tell us anything about how to answer this question.}. We can certainly ask the simulated person whether they feel like they have just woken up from the death bed, but the answer will unequivocally be ``yes'', \textit{regardless} of whether this has actually happened, or whether the actual patient has subjectively died and been replaced by a seemingly identical clone with implanted false memories.

A typical physicalist reaction to this question would be to deny that this is a meaningful distinction to be made in the first place: after all, there is no ``fact of the world'' that could ground the answer this question\footnote{The problem is not whether it would actually be possible to scan the human being and simulate it in a functionally equivalent way --- as mentioned in the first footnote of this paper, here we always make the working assumption that it is. The point is rather that \textit{there is no meaningful question to be asked} from a physicalist third-person perspective, as articulated by the Doctor in Section~\ref{SecPrologue}: the question ``what will probably happen to me?'' is not a question about facts of the world, as its answer is independent of any claims about e.g.\ how matter behaves.}. The disadvantage of this reaction is that it robs us of any hope for guidance if we actually face this situation (like you did in Section~\ref{SecPrologue}), which might well be the case within the next century or so. Furthermore, as the Boltzmann brain (BB) example below will demonstrate, there are situations in physics where we have encountered questions of a similar type (such as ``should I believe that I will make a BB-type observation next?'') whose answers cannot be grounded on facts of the world \textit{either} (even though it may at first seem that they could).

\begin{figure}[t]
\centering 
\includegraphics[width=0.8\columnwidth]{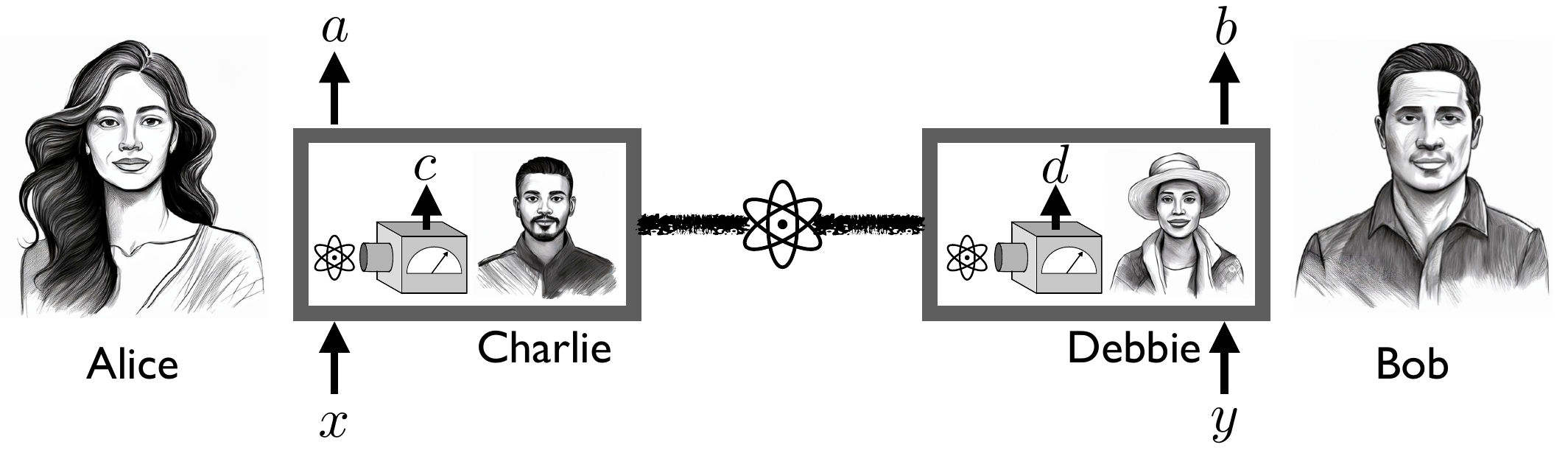}
\caption{Sketch of the thought experiment by Bong et al.~\cite{Bong}. An entangled quantum state is shared between two locations. At one location, Charlie performs a fixed measurement on one of the two particles, obtaining outcome $c$. Alice, however, regards the box with the particle and Charlie as a closed quantum system, and decides to perform a measurement $x$ on it, obtaining outcome $a$. (Note that $x$ is chosen \textit{after} $c$ has been established.) In the special case of $x=1$, she asks Charlie directly for his outcome and outputs $a=c$. At the other location, we have two further observers, Debbie and Bob, who act analogously. If the externally observed statistics $P(a,b|x,y)$ violates a so-called \textit{local friendliness inequality}, then locality and no-superdeterminism imply that there cannot for all $x,y$ exist a joint probability distribution $P(a,b,c,d|x,y)$ that describes the observations of all agents.}
\label{fig_wf}
\end{figure}

Indeed, recent no-go theorems tell us that we \textit{cannot} always ground our beliefs about our own future observations on expected objective properties of the world, unless we give up important principles such as locality\footnote{The locality assumption of Bong et al.~\cite{Bong} is much weaker than the locality assumption in Bell's theorem: it does not refer to hidden variables, but expresses \textit{no-signalling for actually observed outcomes} (``the probability of an observable event is unchanged by conditioning on a space-like-separated free choice $z$, even if it is already conditioned on other events not in the future light-cone of $z$.''). This is necessary to avoid conflict with relativity.}. Consider the thought experiment sketched in Figure~\ref{fig_wf} by Bong et al.~\cite{Bong}, which is a Wigner's friend-type~\cite{Wigner} scenario, but extended to involve four observers and an entangled quantum state. An external observer can repeat the experiment many times, and record the outcomes $a$ and $b$ of the agents Alice and Bob, given their choices of settings $x$ and $y$. Doing so, this observer may determine a relative frequency estimate of the probabilities $P(a,b|x,y)$, and compare the result with the quantum predictions. Assuming empirical adequacy of Quantum Theory, i.e.\ the correct prediction of the outcome probabilities of actually observed events in this scenario, Alice and Bob can use these quantum predictions to determine \textit{what they should believe to observe} in this experiment: Alice should assign the probability
\[
   P(a|x)=P(a|x,y)=\sum_b P(a,b|x,y)
\]
to seeing outcome $a$ if she chooses setting $x$ (the first equality is due to the no-signalling principle, since Alice and Bob are spacelike separated). A similar calculation applies to Bob. If we only consider Alice and Bob, then this resembles a very familiar story as we could also have told it in the context of, say, classical statistical mechanics: Alice can determine the probabilities she should assign to her own future observations by marginalization of the joint probability distribution over all relevant facts of the world at the moment of observation.

In Quantum Theory, we cannot always assign joint probability distributions to all variables that may, counterfactually, potentially be observed in the future, as Bell's theorem demonstrates. However, we could have hoped that there is still a ``classical'' regime of the world where we \textit{can} talk about variables in the usual sense that are jointly distributed, if we restrict our consideration to variables that \textit{are} actually observed. Indeed, this is typically the case in our every-day world: all humans share a common Heisenberg cut, and everything that we observe during our lifetimes can (so far) be shared with all other humans. All observations (outcomes, experiences) are part of a large Boolean algebra of propositions, and we can think of a joint probability distribution that corresponds to what any hypothetical external observer should believe \textit{about the world}. What we ought to believe about our future observations should then be derived from this as a marginal probability distribution, at least in principle (in practice we certainly use heuristic simplifications and shortcuts).

However, this situation will change as soon as we implement extended Wigner's friend experiments such as the one described above in our world. In this case, the results above tell us that the observations $a,b,c,d$ made by Alice, Bob, Charlie and Debbie, respectively, do \textit{not} belong to a Boolean algebra of propositions of the type just described. That is, if the externally observed statistics $P(a,b|x,y)$ violates a so-called \textit{local friendliness inequality}, then there \textit{cannot} for all $x,y$ exist a joint probability distribution $P(a,b,c,d|x,y)$ of the observations of \textit{all four} observers from which $P(a,b|x,y)$ derives, unless either the principle of locality or of no-superdeterminism (or both) are violated. But if there is no such distribution, then the four observers \textit{cannot} all use the ``standard methodology'' to determine what they should expect to see in the experiment. That is, the four observers cannot start with what they believe about the facts of the world (including $a,b,c,d$), and from this deduce their own small part of this (the marginal distribution on $a$ and $b$). Depending on one's interpretation of Quantum Theory, one might conclude that Charlie and Debbie can individually use the Born rule to obtain probabilities $P(c)$ and $P(d)$, but the validity of these assignments can never be verified externally.

In a recent publication~\cite{JonesMueller}, we have introduced some terminology that aims at describing the essential core of this observation:\\

\textbf{Restriction A (informally):} Our physical theories do not (and sometimes \textit{cannot}) give us joint predictions for the future observations obtained by all \textbf{a}gents.\\

More formally, Restriction A is a property of any given theory $T$ (such as Quantum Theory, supplemented by additional assumptions such as locality and no-superdeterminism), and predictions are assumed to be formalized as probability distributions. Restriction A may apply to some scenarios (models, experiments) within $T$, but not others. It applies to $N=4$ observers in the Wigner's friend scenario above, and it applies to a \textit{single} observer, i.e.\ $N=1$, in the simulation scenario of Section~\ref{SecPrologue}.

The cases of $N=1$ and $N\geq 2$ have quite different interpretations. For $N=1$ agent, Restriction A points at a deficiency of the given physical theory: the agent might want to predict its own future observations, but cannot. For $N\geq 2$ agents, Restriction A says that the agent cannot obtain such predictions by predicting properties of the external world that are facts for all agents. We will return later to the question of how to interpret this.

Moreover, suppose that the theory $T$ is \textit{intersubjectively empirically complete} in the following sense: for all experiments for which the results can be recorded and shared among all external observers (not necessarily among all participants in the experiments, such as Charlie and Debbie), the theory $T$ actually supplies us with a statistical prediction for this experiment. If this is the case, then \textbf{Restriction A only applies to theory $T$ in scenarios where the agent's observations \textit{cannot} be externally recorded}; in other words, where there is some sort of ``epistemic horizon''~\cite{Fankhauser2023,Fankhauser2024,Szangolies} that separates some of the observers from some others. This is indeed the case in Bong et al.'s thought experiment: as we have already seen, it is impossible to record the outcomes $a,b,c,d$ in every single run of the experiment, and to obtain statistics that can, say, be published in a scientific journal. The problem is that asking Charlie or Debbie for their outcomes will in general disturb the experiment, and will prevent the violation of a local friendliness inequality. Similarly, in the simulation scenario of Section~\ref{SecPrologue}, it was impossible for any third person (such as the Doctor) to even \textit{decide whether} the event that the agent was wondering about has actually happened or not in any single implementation. After all, this was the reason for the Doctor to dismiss the patient's question: a physicalist might want to claim that the third-person facts of the world are all there is to say, and the main characteristic of a ``fact'' is its intersubjective external accessibility.

In Ref.~\cite{JonesMueller}, we have explained in detail why we think that Restriction A is at the core of several other enigmas in the Foundations of Physics and Philosophy. Here I only briefly sketch two of them. The first puzzle is cosmology's \textbf{Boltzmann brain problem}. For a thorough introduction, I refer the reader to Carroll's work~\cite{Carroll}; the following is an extremely brief and cartoonish summary. Suppose that we have a model of the universe that predicts it to be combinatorially large. In this case, thermodynamic fluctuations will produce all kinds of unlikely events in this universe, including Boltzmann brains: local collections of matter that, by mere chance, are functionally equivalent to a human brain, complete with false memories and thoughts such as ``I have lived on Earth for 30 years''. Now, what if our cosmological model predicts that there are many more (of the order, say, $10^{100}$) Boltzmann brains than ``ordinary'' brains that have evolved on planets? An intuitive thought is that, in this case, we should expect to be Boltzmann brains. But Boltzmann brains would typically make some very unexpected observations soon (if they don't disintegrate right away), such as seeing high-temperature radiation instead of the usual night sky. Since this is not what we observe (and we do not seem to disintegrate either), we seem to not be Boltzmann brains\footnote{As pointed out by Carroll~\cite{Carroll}, a more thorough and careful argumentation should revolve around the notion of \textit{cognitive instability}: if a physical theory gives us reasons to believe that we are Boltzmann brains, then this theory undermines all reasons to trust its predictions in the first place.}. Does this mean that we have just falsified the corresponding cosmological model?

It is important to note that cosmology, or our current physical theories more generally, only tell us (an estimate of) the number of Boltzmann brains and, perhaps, of ordinary brains, given some cosmological model. But they do \textit{not} tell us  \textit{what you should believe} if you wonder whether you are one or the other --- in particular, which probability you should assign to making an extraordinary Boltzmann-brain-type observation next. For example, in order to claim that counting frequencies should give you those probabilities amounts to accepting something along the lines of Elga's Principle of Indifference~\cite{Elga}, which represents a logically independent \textit{addition} to our physical theories\footnote{In~\cite{JonesMueller}, we argue that this principle is not as well-motivated as it might first seem, and that there are other plausible choices too, including ones that would lead to very different conclusions from counting.}. Clearly, it would be beneficial for cosmologists to be able to use this reasoning to rule out Boltzmann-brain-dominated models, but Restriction A for $N=1$ observer applies: physics does not tell you whether you should believe that you are a Boltzmann brain.

A second class of puzzles where Restriction A applies are situations that resemble Parfit's teletransportation paradox~\cite{Parfit}: these are scenarios where a given observer is undergoing some sort of duplication procedure (think of the Star Trek episode ``The Enemy Within''). For example, suppose that you will be scanned in all detail (while destroying the original, as in Section~\ref{SecPrologue}), and a large number $N$ of copies will be created which will all make different observations in the future. What should you believe about your future experiences in this case? At first sight, it seems like assigning equal probability to every copy is natural, but what would you do if, in fact, an infinite number of copies were created, and with increasing $n$, the $n$th copy will more and more deviate from the original, but only ever so slightly from copy $n$ to $n+1$? What if the copies are created also at different \textit{times}? We have no idea what to say in this case, and our physical theories have nothing to say about what you should believe to experience next in this scenario\footnote{I do not believe that it is relevant to acknowledge that we will not \textit{practically} run into a scenario of this form any time soon: our way to respond to those puzzles should in principle apply to \textit{all logically conceivable} situations; if it does not, this is a sign of incomplete understanding. Furthermore, Everettians in particular might want to acknowledge that this scenario is perhaps not that extraordinary after all.}. This is an instance of Restriction A for $N=1$ observers. In~\cite{JonesMueller}, we construct a more elaborate version of a multiplication thought experiment that demonstrates an instance of Restriction A for $N=2$ observers, robust to future changes of our physical theories, which resembles a probabilistic version of the Frauchiger-Renner paradox~\cite{FrauchigerRenner}.

Algorithmic idealism, as described below, can be understood as a reaction to Restriction A in the following sense. In this conceptual framework, Restriction A for any given \textit{single} agent ($N=1$) is seen as a serious problem that needs to be alleviated, because it reveals the inability to predict something about the results of \textit{actual experiments that somebody can do}. For example, \textit{you} can in principle (and, in the perhaps not so far future, in practice) agree on being copied into a computer simulation in some sense, \textit{and see what happens to you}. Despite the fact that the outcome cannot be intersubjectively shared (as explained above, this is a prerequisite for Restriction A to even apply in the first place), it seems absurd to believe that there was nothing that could be said (probabilistically, say, or plausibilistically~\cite{FritzLeifer}, or in some other sort of mathematical formulation) about which sort of outcome of this experiment you ought to expect to experience. After all, the world will kick back at you and generate some subjective outcome, and something must determine the nature of this outcome. Even claiming that \textit{nothing can be said} in this case is just an utterance of a human being in need of clarification, which means in need of mathematical formalization. Indeed, algorithmic idealism suggests a particular resolution of Restriction A for $N=1$ agents via Postulate 2 below (and via its formalization in any specifically chosen mathematical model).

However, algorithmic idealism does not claim to resolve Restriction A for $N\geq 2$ agents, but rather suggests to accept it: according to its predictions, different agents \textit{do not in general} share a common world into which they seem to be embedded. Instead, ``objective reality'' is an approximate and emergent statistical phenomenon that does not apply in all cases (see Subsection~\ref{SubsecReality} below). Indeed, Restriction A for $N\geq 2$ observers (and the recent results~\cite{Bong} showing that, under some natural assumptions, all future physical theories will suffer from it) is taken as corroborating evidence for the correctness of algorithmic idealism's strategy to drop the assumption that agents are fundamentally embedded into external worlds.

\section{Algorithmic idealism in a nutshell}
\label{SecAlgorithmicIdealism}

How do we respond to Restriction A? In the case of $N\geq 2$ agents, the nonexistence of a joint probability distribution for the observations of all agents can be interpreted\footnote{It certainly does not \textit{have} to be interpreted in this way (e.g., one might drop some background assumptions such as locality that underly the corresponding no-go theorems), but it is a well-motivated and natural possibility.} as evidence that we should not always regard observations or events as absolute (see e.g.~\cite{Fuchs2017,DeBrota,Zwirn}), and algorithmic idealism as described below will agree with this. The case of Restriction A for a \textit{single} agent, $N=1$, is however even more troublesome: it means that an agent can perform an experiment where they will learn some outcome, but there is nothing that can be said about what they should expect to observe. As explained further above, I regard this as absurd, and interpret it as a shortcoming of our given theories. If one takes this viewpoint, how does one react to it?

One option is certainly to double down on physicalism, and to insist on saying that ``what should I expect to experience next?'' is not usually a well-defined question. In this view, we should see this as a kind of pragmatic question that sometimes does not have an answer, and that will often have a \textit{vague} answer, similarly as the question \textit{``Are viruses alive?''}

However, here I suggest to explore a second option, which is to assume that (a specific version of) this first-person question \textit{does} always have an objective answer, and to work out a theory that aims at supplying this answer at least in principle. The motivation to do so is two-fold: pragmatically, we will at some point be in desperate need of guidance in exotic situations such as the one of Section~\ref{SecPrologue}, and a precise theory that offers such guidance and that is \textit{also} in agreement with the standard physics predictions in mundane situations seems to be a good tool to aim for. Furthermore, as I have argued in Section~\ref{SecQuantum}, Quantum Theory suggests that the question of ``what should I expect to observe next?'' is a more fundamental, natural and fruitful question to ask than ``what is the case in the world''. We may see this as a hint from modern physics that we can hope to get closer to the truth by attempting to answer the former rather than the latter, and we can hope that trying to do so may indirectly tell us something interesting about the world, too.

The goal of algorithmic idealism is to implement this second option. Despite its name, I am aiming for a formal, mathematically rigorous approach for which higher-level concepts such as consciousness or belief do not play any fundamental role.

\begin{figure}[h]
\centering 
\includegraphics[width=0.8\columnwidth]{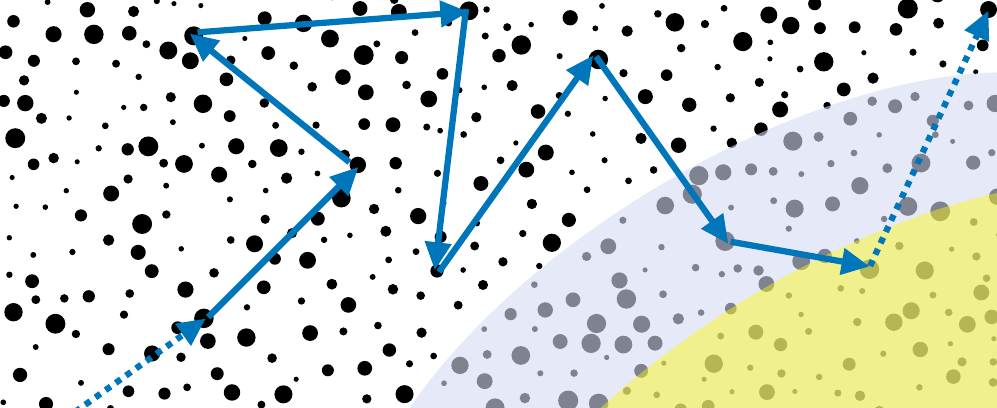}
\caption{The set of self states $\mathcal{S}$, which will typically be modelled as a countably-infinite set, illustrated as a set of black dots. Self states are simply abstract patterns (perhaps formalized as finite binary strings or natural numbers), and only a small subset of those (indicated in light blue) will have an interpretation as describing a (momentary snapshot of an) \textit{agent}. Among those, a very small subset (highlighted in yellow) will have an interpretation as describing \textit{conscious} self states. Neither agenthood nor consciousness, but only the abstract mathematical description of this set is relevant for algorithmic idealism, and it will in particular contain a structure that says which next self states $y$ are more likely, given a current self state $x$. If this is mathematically described by a conditional probability distribution, then we obtain a Markov process; however, even in the simplest model of algorithmic idealism (the bit model), this likelihood is described by an infinite collection of algorithmic priors. Some self states will be more likely to obtain on average than others, hence the different dot sizes. What is not shown is the abstract \textit{computability structure} which will in the end determine the likelihood of transitions, corresponding to what a universal method of induction would predict. {\tiny[Thanks to www.freepik.com for the black dot picture.]}}
\label{fig_selfstates}
\end{figure}

A first step is to put aside the concept of ``belief'' and instead to focus on the current state of the observer (for any given single, subjective moment), in an abstract, information-theoretical sense. For a human observer, for example, this could be the pattern~\cite{DennettPatterns} of neuronal connections in their brain (this description is only meant to be helpful for our intuition and not to become an actual part of the theory). This corresponds roughly to what Parfit called a ``person stage''~\cite{Parfit}. Among other things, this implies that ``agents'' or ``observers'' will \textit{not} be primitive notions of algorithmic idealism, but only momentary ``snapshots'' of the pattern, structure or information content that defines an agent or observer \textit{at some given moment}. These will be called ``self states''\footnote{In the earlier publication~\cite{Mueller2020}, these have been called ``observer states''.}, see Figure~\ref{fig_selfstates} for an illustration and further comments.

A second step is to let go of all human-centric aspects and to pursue a completely abstract approach: a self state can be \textit{any} sort of pattern, in some mathematically, abstractly defined set of all possible patterns. Only a tiny subset of those patterns would deserve to be interpreted as describing something like the momentary state of an ``observer'' (let alone a human or conscious observer). At the level of fundamentality that we are targeting, it would be completely off the mark to introduce a specific model of an observer, or to make any specific assumptions about its composition or function. The most modest and thorough approach is a maximally abstract one, in which all aspects of how we think of observers or agents are tentatively treated as contingent.

This leads us to a formulation of the first postulate of algorithmic idealism:\\

\textbf{Postulate 1} (Self States). \textit{There is a set $\mathcal{S}$ of ``self states'', the collection of patterns that one can be at any given subjective moment. Everything that is to be said about a given agent at some moment, including all inferences to be drawn about its past or future, is determined by its self state.}\\

In particular, and perhaps most counterintuitively, algorithmic idealism \textit{denies that agents are fundamentally embedded into some external world}: an agent at some given moment is \textit{completely} characterized by its self state (which is a standalone pattern), and \textit{not} by its spatiotemporal location inside some external world. In cases such as the Boltzmann brain problem, for example, where there are many locally indistinguishable copies of an agent, it rejects the usual ``self-locating uncertainty'' intuition that the agent is \textit{actually} one of the copies, but does not know which one. Instead, according to algorithmic idealism, the agent (at that given subjective moment) is its self state, and this abstract pattern just happens to be represented, or realized, in each one of the copies. In some sense, the agent should think of itself as the \textit{equivalence class} of all these realizations (ordinary brains and Boltzmann brains and all other, potentially completely different, realizations).

At first sight, this seems absurd: after all, will the agent not make another observation soon that will either confirm or disprove its concerns of being a Boltzmann brain, and hence retrospectively say that it \textit{was actually} the copy on the planet, and not the random fluctuation out there? On second thought, however, the apparent contradiction gets resolved by two insights. First, algorithmic idealism is constructed to avoid the ``standard methodology'' described in Section~\ref{SecRestrictionA}, and to allow agents to predict what they will see next \textit{without} grounding this on the expected state of the world into which they are embedded. Hence, by construction, what they should believe \textit{cannot} depend on whether they are ``actually'' surrounded by a planet-like environment or high-entropy radiation. Second, as will be explained in Section~\ref{SecPredictions}, algorithmic idealism predicts that agents can often think of themselves as ``effectively embedded'': what happens to them will often look to them to excellent approximation \textit{as if} they were a part of some external world with algorithmically simple, probabilistic laws of nature. In the cosmology example, this results in the consequence that the agent will probably experience ``planetary business as usual'' next, regardless of the number of Boltzmann brains, even though the question of whether it \textit{actually is} a Boltzmann brain or not is ill-defined. This will be discussed in more detail in Section~\ref{SecPredictions} below.

Dropping the standard methodology, we now need a postulate that implies what an agent\footnote{I have remarked above that the notion of ``agent'' is not a primitive of algorithmic idealism, but only the notion of a ``momentary snapshot'' in terms of a self state. However, it will still be \textit{linguistically} convenient to use the words ``observer'' or ``agent'' in the colloquial descriptions, simply because it would otherwise be difficult to convey the right intuitions for us human beings who are trained to think in terms of things that persist over time.} should believe to experience next, and since we are not grounding this on an external world in the usual sense, the answer can only depend on the agent's momentary self state. Since we have decided to do without vague terms such as ``belief'' or ``experience'' in algorithmic idealism, this question has to be replaced by another one: \textit{If an agent is in self state $x$ now, in which self state $y$ will it probably be next?} By construction, we cannot rely on our usual laws of physics to ground this question. Instead, a fundamental idea of algorithmic idealism is to rely on a weaker assumption that underlies our trust in empirical science in the first place: that \textit{induction is possible}. More specifically, I implement this idea in the form of the following postulate:\\

\textbf{Postulate 2} (State Change). If you are in self state $x$ now, you will be in another self state $y$ next, and which one this will be manifests itself as a random experiment for you. The likelihood of some $y$ next, given $x$ now, has an \textit{objective} value, which expresses what you \textit{should} believe will happen to you next, if you knew your current self state. The mathematical structure that defines its value formalizes a universal method $M$ of induction.\\

Induction, broadly speaking, is a method to predict future observations by extrapolating regularities of past observations. What I mean by ``universal'' is a method that does not only work under the assumptions that we navigate within our own universe, but in a large class of conceivable environments. The possibility of inductive inference is one of the main assumptions of the scientific method. Here, we stipulate that induction should in principle \textit{always} be possible, even in exotic situations, but in the following \textit{indirect} way: there is a method of induction $M$ such that the likelihood of becoming $y$ next, given that one is $x$ now, is exactly the one that is predicted by $M$. In other words, a hypothetical third person that uses method $M$ could correctly (albeit not deterministically) predict any other agent's future self state, if it were given a complete description of that agent's current self state. A consequence of Postulate 2 is that \textit{regularities that are present in your current self state will tend to persist in your private future}, because methods of induction assign high probability to this being true.

Probabilities obtained by some method of induction are naturally interpreted in a \textit{Bayesian} way: many such methods begin with a certain choice of prior, and apply Bayesian updating upon obtaining new information (this is in particular true of Solomonoff Induction, used in the bit model below). However, here, the mathematical formulation of a method of induction is detached from its epistemic origins: rather than describing someone's beliefs about an unknown future, the values that it defines mathematically are now interpreted as some sort of \textit{objective chance} of turning from some given self state to another one. In more detail, they are supposed to be interpreted in a similar way as Berghofer's DEJI interprets quantum probabilities: not as expressions of what somebody believes, but what they \textit{should} believe if they were told the current self state\footnote{This does \textit{not} imply that the \textit{agent itself} (which is in self state $x$ now) can use method $M$ to predict its own future experiences directly. After all, the agent will not in general have complete knowledge of $x$, and some self states will represent observers that are incapable of reasoning, or that are simply patterns which defy any interpretation as ``observer'' or ``agent''. However, agents like us who \textit{are} able to reason rationally may still use the resulting model of algorithmic idealism to predict something about their future observations. This can be done either by working out \textit{general} predictions of algorithmic idealism that are independent of the specific form of the current self state $x$, or by analyzing the consequences of \textit{some} (incomplete, but non-zero) knowledge of $x$.}.

Postulates 1 and 2 are intentionally formulated in such a way that they admit of several possible rigorous mathematical formulations, or \textit{models}. For every choice of self states $\mathcal{S}$ and method of induction $M$, one obtains a different model. One simple, preliminary model, the \textit{bit model}, has been introduced in~\cite{Mueller2020}. It will be described in Section~\ref{SecAIT} below. Different mathematical models (and the associated methods of induction) may rely on different mathematical definitions of likelihood, e.g.\ on a single conditional probability distribution, on a collection of \textit{several} such distributions (which is the case for the bit model below), on a notion of imprecise probability~\cite{Walley}, or on plausibility measures~\cite{FritzLeifer,Friedman}. The precise conceptual interpretation of these ``values'' of likelihood, as currently sketched in Postulate 2, may have to be adapted on a case-by-case basis. Algorithmic idealism should therefore be seen as a kind of \textit{conceptual framework} rather than a single theory, because it admits many different mathematical realizations.

As an analogy, consider the general idea that physical space could be curved rather than flat, or more generally, non-Euclidean in some way. This single idea has many possible mathematical realizations that make very different physical predictions, even though \textit{some} predictions may be universal (such as a violation of the triangle postulate). The question of which one of the resulting theories is correct (if any) is ultimately a matter of empirical observation and experiment. In fact, the idea that non-Euclidean geometry applies to our world predates the theory of relativity. In 1872, the German astrophysicist Karl Friedrich Z\"ollner~\cite{Kragh} suggested that we may live in a non-Euclidean curved universe. This allowed him to suggest a solution to what is known as Olbers' paradox: if the universe had positive curvature, it could have finite volume without having a boundary, which would allow us to understand why the night sky is dark. Ultimately, further experimental results, such as the perihelion precession of Mercury's orbit or the outcome of the Michelson-Morley experiment, were necessary to arrive at yet another modification of the conceptual framework which ultimately resulted in General Relativity: that we should consider curved \textit{spacetime} rather than space. Nevertheless, it would be fair to say that Z\"ollner's idea turned out to be essentially correct: finiteness of a non-Euclidean universe resolves Olbers' paradox.

In hindsight, while Z\"ollner should not have hoped to guess the correct theory of the universe a priori, he would have been correct to anticipate the general field of mathematics that would be involved in its formulation: differential geometry, because this is the arena of reasoning that allows us to describe spatial geometries which are locally familiar but potentially globally unusual. We can similarly guess what field of mathematics will be relevant for all formalizations of algorithmic idealism. In the \textit{standard} picture of an agent or observer, we would say that the observer's state contains a large amount of memory and experiences \textit{of an external world}; that is, variables which are \textit{correlated} with the variables that comprise relevant aspects of the outside universe. This is the paradigmatic scenario described by (standard) information theory~\cite{CoverThomas}: essentially, an observer's state would correspond to information, and it would be \textit{information about} an external world. However, algorithmic idealism concerns the structure of the observer (at some given moment) and its regularities (because a method of induction must be applicable) \textit{without} reference to anything external. We need an area of mathematics that is concerned with the structural regularities in a bare, concrete collection of standalone data, not in a random variable that is potentially correlated with another variable. Such a field exists: algorithmic information theory (AIT)~\cite{Hutter,LiVitanyi}, and hence the name of the approach.

\section{Algorithmic information theory and the bit model}
\label{SecAIT}
AIT is concerned with the information content of individual chunks of data, usually described in terms of the finite binary strings, $\{0,1\}^*=\{\varepsilon,0,1,00,01,10,11,000,001,\ldots\}$, where $\varepsilon$ is the empty string, or the integers, $\mathbb{N}_0=\{0,1,2,3,4,5,\ldots\}$. The most well-known quantity studied in AIT is \textit{Kolmogorov complexity}: if $x$ is a binary string or an integer, then $K(x)$ denotes the length of the shortest program for a universal Turing machine $U$ that produces $x$ as its output. In the following, let us focus on the binary strings for convenience. Suppose that we generate a string $x$ of length $n$ by tossing a fair coin $n$ times. Then, with high probability, $K(x)\approx n$, because the shortest computer program to generate $x$ will typically be ``print $x$'', i.e.\ a program that lists all the bits of $x$. On the other hand, if $x$ is a string of $n$ zeros, $x=\underbrace{000\ldots 0}_n$, then $K(x)\approx \log n$, because we only have to tell the Turing machine how many times it has to print a zero, which can be done by specifying the approximately $\log n$ (in base $2$) binary digits of $n$. Similar reasoning applies to the first $n$ binary digits of $\pi$, or of $\exp(\sqrt{2})$, for instance.

The exact definition of Kolmogorov complexity depends on some choices of the Turing machine model that is used in its construction. For example, Kolmogorov complexity $K$ defined above is usually defined for so-called \textit{prefix-free} Turing machines (for which every program needs to describe where it ends), but there is another version called $C(x)$ for ``bare'' machines without this requirement. The fact that there are many possible definitions of its basic quantities implies many different possibilities to potentially ground a formalization of algorithmic idealism on AIT.

AIT also allows us to ask, for example, how likely it is that the Turing machine $U$ generates some given output $x$, if it is supplied with random input. How this is defined exactly depends on choices about the details of the working of the Turing machine. For prefix-free Turing machines, the textbooks give the definition of \textit{algorithmic probability}
\[
   m(x):=\sum_{p:U(p)=x} 2^{-\ell(p)},
\]
where $\ell(p)$ denotes the length of the binary string (program) $p$. This is interpreted as the probability that $U$ outputs $x$ on input that is chosen by a sequence of fair coin tosses. It turns out that
\begin{equation}
    -\log m(x)=K(x)+\mathcal{O}(1).
    \label{eqm}
\end{equation}
 This notation means that the difference between $\log m(x)$ and $K(x)$ is bounded in absolute value by a global constant that is independent of $x$. This example already demonstrates some features of the scientific practice of AIT: researchers in this field are typically not interested in the actual \textit{values} of quantities like $K(x)$ or $m(x)$ (in fact, these numbers are typically not computable), but in the scaling of these quantities with respect to other quantities such as the length of $x$, or the comparison of such quantities\footnote{In some sense, this resembles scientific practice in physics, where we sometimes say things like ``Suppose that the interaction energy $\Delta$ is small compared to the total energy $E$'' etc.}. This attitude also eliminates most of the worries arising from the freedom of choice of universal Turing machine $U$: for any other universal machine $V$, we would have $K_U(x)=K_V(x)+\mathcal{O}(1)$, where $K_U$ denotes Kolmogorov complexity as measured with respect to the machine $U$. That this sort of curious interplay between exact definition and (sort of) vague comparison-type application leads to a meaningful field of inquiry (and, in fact, to immense mathematical beauty) is a fact that cannot be conveyed in terms of the theory itself, but only be experienced when actually working with AIT.

Equation~(\ref{eqm}) says that strings $x$ which are \textit{simpler} (in the sense of having a shorter description on a universal Turing machine) are \textit{more likely} (with respect to the distribution $m$). This is a general phenomenon for probabilistic quantities introduced and studied in AIT: they favor simplicity, resembling the intuition of Occam's razor. Crucially, AIT allows us to define conditional versions of $K$ and $m$ that are potentially relevant for formulations of algorithmic idealism. For example, expressions\footnote{The definitions of these quantities are not relevant for the present discussion, and they can be found in~\cite{LiVitanyi}.} such as $K(y|x)$ or $C(y|x)$ quantify how much additional information a machine needs to produce $y$ if it is given $x$ for free; and conditional probability distributions such as $m(y|x)$ or $M(y|x)$ tell us how likely it is that a machine will output $y$ if it is given $x$ for free, resp.\ how likely it is that the machine will output $y$ after it has output $x$.

Intuitively, conditional probabilities such as $M(y|x)$ will be large if and only if it is easy to generate $y$ from $x$ on a Turing machine, e.g.\ if $y$ contains many of the computable regularities of $x$. If so, then this suggests the idea that an expression like $M(y|x)$ might be be used as the basis of a method of inductive inference, and then perhaps the result could play the role of the method of induction mentioned in algorithmic idealism's Postulate 2 above. There is at least one version of conditional probability for which this intuition can be made rigorous, resulting in a method of inductive inference that is known as Solomonoff induction~\cite{LiVitanyi}:\\

\textbf{Solomonoff induction.} \textit{Suppose that you observe one bit after the other, and you have seen the $n$ bits $x_1^n=x_1 x_2 \ldots x_n$ so far. Then, predict that the next bit will be $b$ with probability $\mathbf{M}(b|x_1^n)$, which is the probability that the universal monotone Turing machine (defined in~\cite{LiVitanyi}) will output $b$ after having output $x_1^n$.}\\

That this is a good method of prediction can be shown as follows~\cite[Corollary 5.2.1]{LiVitanyi}. Suppose that the bits that you see are \textit{actually} generated by some computable process $\mu$ (a ``recursive measure'' in the terminology of~\cite{LiVitanyi}), which you do not know. Then, with $\mu$-probability one,
\begin{equation}
   \lim_{n\to\infty} \left| \mathbf{M}(b|x_1^n) - \mu(b|x_1^n)\right| = 0\qquad (b\in\{0,1\}).
   \label{eqSolInd}
\end{equation}
That is, in the long run, algorithmic probability $\mathbf{M}$ will almost surely agree with the actual probabilities $\mu$ determined by the unknown computable process that generates the bits that you see. For example, suppose that the process always outputs deterministically the bit $1$, i.e.\ $\mu(1^n)=1$ for all $n\in\mathbb{N}$, where $1^n=\underbrace{11\ldots 1}_n$. Then $\mathbf{M}(0|1^n)=2^{-K(n)+\mathcal{O}(1)}$, which tends to zero and is of the order $1/n$ for most $n$. In this sense, after seeing $n$ ones, Solomonoff induction predicts that you will probably not see a zero next, as expected from a method of induction.

Based on this result, I have constructed a preliminary formalization of algorithmic idealism in~\cite{Mueller2020}, the \textbf{bit model}. Its set of self states is the set of finite binary strings, $\mathcal{S}=\{0,1\}^*$. The method of induction is Solomonoff induction. More concretely, the claim is that self state $x$ is always followed either by self state $x0$ or by self state $x1$ (that is, $y$ is obtained by appending a single bit $b$ to $x$) and the probability is given by $\mathbf{M}_U(b|x)$. In more detail, the likelihood of changing self state to $y$ is characterized by the set of all $\mathbf{M}_U(b|x)$, over all universal monotone Turing machines $U$, and the predictions of the theory are by definition those that are invariant under the choice of $U$. That is, the notion of ``likelihood'' is a more liberal one than a specification of a fixed probability distribution, and this is discussed in detail in~\cite{Mueller2020}.

This extremely simple model is quite powerful: it admits the rigorous proof of several significant predictions that I will describe in the following section. However, it has some undesirable features that motivate my current work on an improved version. In particular, it implies that a momentary state $x$ has always complete information about what the \textit{previous} self states have been (namely, the prefixes of $x$). This is both unrealistic (thinking, say, of human observers) and unmotivated. It is simply an artefact of the way that AIT has been pursued so far.

In the rest of the paper, I will focus on the conceptual aspects of the theory and set all technical details of AIT aside. In particular, instead of giving technical proofs of its main predictions in a particular realization such as the bit model, I will describe the main ideas for why these predictions should be expected to follow from Postulates 1 and 2.\\

\section{Predictions of algorithmic idealism}
\label{SecPredictions}
Every novel theory must reproduce the predictions of earlier theories to good approximation, in agreement with existing empirical results, and then it must do more. For algorithmic idealism to be successful, it should
\begin{compactitem}
\item be consistent with the predictions of our physical theories in standard situations;
\item explain facts that are simply assumed in our physical theories, e.g.\ why it typically appears to us that we are embedded into a probabilistic world with simple laws of nature; and
\item give us concrete (and perhaps surprising) predictions for exotic scenarios such as the Boltzmann brain problem or the computer simulation of agents.
\end{compactitem}
In this section, I will argue that algorithmic idealism can satisfy all these desiderata. I will do so via plausibility arguments, based on Postulates 1 and 2 of Section~\ref{SecAlgorithmicIdealism}. In~\cite{Mueller2020}, I have given rigorous formulations and proofs of these claims for the bit model as described above. However, for the reasons mentioned above, the bit model has to be improved. My hope is that the colloquial argumentation of the present paper can be a useful guide for the proof of the claims in other mathematical formalizations of algorithmic idealism. It certainly gives a valid conceptual summary of what is going on in the technical proofs for the bit model of~\cite{Mueller2020}.

\subsection{Consistency with physical predictions in standard scenarios}

The main argument for why algorithmic idealism will be in agreement with our physical theories is that \textit{physics seems to predict that induction is possible}. Consider an agent characterized by some self state $x$ which contains a lot of data. As we have said above, a self state need not correspond to anything that we typically interpret as an agent or observer, so let us consider planet Earth. Motivated by AIT, we may focus on versions of algorithmic idealism where the set of all possible self states is the finite binary strings, $\mathcal{S}=\{\varepsilon,0,1,00,01,\ldots\}$. To relate Earth with a binary string, let us fix an arbitrary method for encoding an approximate description of a momentary\footnote{We may have to decide how to deal with relativity's lack of absolute simultaneity and other ambiguities, but any decision for how to do so will considered to be part of the definition of the encoding procedure. {Mathematically, the procedure must be computable, see the notion of a \textit{computational ontological model} of the technical paper~\cite{Mueller2020}.}} configuration of Earth into bits. For example, we might coarsegrain the shell of the sea level plus or minus one kilometer into approximate cubes of sidelength $10^{-9}$ meters, and store a couple of bits for each cube that describe the approximate material composition of the corresponding cube, resulting in an enormous chunk of data.

For the sake of the argument, imagine that we give this data to a team of brillant scientists that has access to astronomical computational power. We ask the team to make a prediction for what this source of data will yield next. (Even if we think of continuous time evolution in our world, the result of this discrete encoding procedure will \textit{discretely} change in a non-zero but very small fraction of a second in the future, and this change is what the third party is supposed to predict.) Earth contains a \textit{lot} of records about the universe in which it is located: there are fossils in the underground implying biological evolution, astronomical photographs in libraries that convey some knowledge about what the world looks like external to Earth, and movies or other subtle properties of its structure that reveal many aspects of the physical laws in the relevant regime. It is not unreasonable to expect that a \textit{complete} description of the relevant laws of physics and, say, the relevant aspects of the state of the universe at the Big Bang can be extracted from this data in principle -- after all, this is what we humans seem to accomplish successfully via science. Hence, the team of scientists should be able to accomplish their task successfully.

What this thought experiment then shows is the first step in the following three-step argument:
\begin{enumerate}
\item We have heuristically argued, given the success of science, that good methods of induction -- if they were supplied with a description of a momentary state of Earth as just described -- should predict the same probabilities of future states of Earth as our best scientific theories. Hence if science makes correct predictions, good methods of induction will do so as well.
\item As a next step, we lift this from a pragmatic, heuristic, methodological intuition to a precise mathematical claim: if we are given a \textit{mathematically well-defined} method of universal induction $M$ (such as Solomonoff induction defined above), then $M$ should \textit{also} make correct predictions about the probabilities of future states of Earth. This is much less intuitive, since the input to $M$ is simply a bare self state; in the bit model, it is a long string of binary data. It may seem surprising (and for philosophically-minded readers highly dubious) that it is enough to have this data \textit{without any built-in semantics}, such as a reference to spatiotemporal structure etc., and \textit{still} to obtain correct predictions from it. But this is guaranteed by mathematical theorems (essentially eq.~(\ref{eqSolInd})) if the induced stochastic process on the self states is \textit{computable}.

Indeed, the laws of physics as we know them are computable in some sense, which is described in more detail in physical versions of the Church-Turing thesis~\cite{Deutsch,Wolfram,Gandy,ArrighiDowek,Hofstadter,Piccinini}, see also~\cite[p.\ 11, footnote 8]{Mueller2020}. And this seems to imply that finite-data processes such as the one described above (discretized Earth) are computable in the right technical sense to conclude that Solomonoff induction\footnote{To apply the bit model, and hence Solomonoff induction in its standard formulation, we would have to set up the encoding such that the binary string can only grow over time by appending new bits. Constructing a model that does not suffer from this restriction is the subject of ongoing work.} $M$ makes asymptotically correct predictions.
\item Steps 1 and 2 have only argued about general properties of universal induction, formulated via well-known results of algorithmic information theory. Algorithmic idealism becomes only relevant \textit{now}: it says that \textit{if you are Earth's current self state}\footnote{As discussed further above, self states are not in general associated to intuitive instances of ``agents'' or ``observers''.}, \textit{then what you will become next is determined by $M$ (in the bit model, algorithmic probability).} In particular, it is \textit{not} determined by the laws of any world in which you might imagine being embedded (actually, you are unembedded). Instead, $M$ is the ``law of nature'', and its action depends on the bare self state and not on anything external to it.

What steps 1 and 2 now show is that algorithmic idealism's claim of \textit{what happens to you if you are Earth (in more detail, if you are its data defined by the specific choice of read-out procedure)} is consistent with \textit{what happens to Earth according to our usual physical picture of the world}.
\end{enumerate}

In the above, there was nothing special about Earth or about the specific choice of encoding procedure. We could instead think of \textit{anything else} that contains a large amount of data about our world, for example a human brain, a hard disc, or the left half of the brain of some mammal, to some level of detail of description. In all those cases, Postulate 2 should yield very similar (and asymptotically identical) claims for \textit{what happens to you if you are that thing\footnote{To be more precise, it is not about being ``that thing'', but about being the self state that is encoded into it.}} as physics does.

In the bit model, we can be more specific what the requirements are for this consistency result to hold. Consider some bit-string-valued random variable $x$ in the world (see the technical paper~\cite[Example A.1]{Mueller2020} for how this is consistent with quantum theory), changing over time in a way that its length does not decrease. For example, think of a video camera that is continuously filming Times Square and recording the pictures on a hard disc, where the video is encoded into bits. Or think of the ``state of the Earth'' example above, or a discretization (and arbitrary computable method of digitization) of your brain to some large accuracy over time, keeping all past information. Our physical theories tell us that the physical world generates a probability measure $\mu$ on the possible histories of this bit string (for example, pictures of the Times Square showing all pedestrians suddenly floating up towards the sky has extremely small probability). If a suitable version of a physical Church-Turing thesis is true, then $\mu$ will be a computable measure. (For this, not only the process defining the relevant aspects of the world, but also the definition of the random variable needs to be computable.) Then, Eq.~(\ref{eqSolInd}) will be true: predictions via algorithmic idealism will in the long run agree with the predictions via physics.

Recall that Postulate 2 is not a description of what we (or the agent) \textit{believes} will happen next, but what they \textit{should believe}. That is, it is understood, in its rigorous formulation within a specific choice of mathematical model (such as the bit model), as a statistical law of nature. In particular, it is not means as a \textit{supplement} of the given laws of physics, but as a \textit{replacement}\footnote{It would certainly be a very bad idea to try and use algorithmic idealism, or algorithmic probability, directly in the regime of physics. It will never be a good replacement \textit{in practice}, it is only a very abstract theoretical (but important) fact that it could be \textit{in principle}.}. Therefore, it is crucial to have a theorem that guarantees its consistency with the standard physical laws in the regime where they apply. This is essentially the regime of empirical science, i.e.\ of processes where a large amount of data is collected over time.

\subsection{The external world as an emergent approximate phenomenon}

Algorithmic idealism predicts this sort of consistency with physics, but all the above reasoning says is that agents who appear to be embedded in a simple probabilistic universe are also likely to make observations in the future that confirm this. But \textit{why should this former premise be the case in the first place}, if algorithmic idealism is true? After all, one of the main methodological principles of algorithmic idealism is to drop the assumption that an agent is fundamentally embedded into an external world. So how can we understand why agents like us are \textit{effectively embedded}, i.e.\ that what happens to us appears \textit{as if} we were embedded?

One main insight shown for the bit model, but hopefully true much more generally, is the following somewhat technical result within AIT. Algorithmic probability measures such as $\mathbf{M}$ are not computable. However, they may often behave very similarly as a computable measure $\mu$, in particular if $\mu$ is algorithmically simple. This may in particular be the case if, by chance, the previously observed bits $x_1^n=x_1 x_2 \ldots x_n$ happen to correspond to a typical result of the probability measure $\mu$; intuitively, in this case, the following bits will \textit{also} likely be distributed in accordance with $\mu$. Intuitively, the statistical regularities present in $x_1^n$ will tend to be preserved because this is what induction deems likely to be the case. This leads to a general tendency of ``stabilizing'' a very regular statistical behavior: one that is according to a simple computable measure $\mu$. If this turns out to be the case, then the evolution of the agent's self state will approximately be described by $\mu$, and there will \textit{by definition} exist\footnote{This is a statement of \textit{mathematical} existence, not a claim of actual physical implementation, whatever the latter would mean. In particular, algorithmic idealism does \textit{not} claim that there are actual machines that perform the computations which will naturally arise as external worlds for our experiences; these computations are in some sense extremely convenient (and hence meaningful) fictions.} a computation, based on a short program that generates $\mu$. To predict what happens to the agent, we may hence \textit{pretend that the agent is part of a computational process that generates its self state according to the computable distribution $\mu$}, and this process plays the role of an external world.

In the case of the bit model, the technical result that I mentioned can be described as follows (see~\cite{Mueller2020} for the details):\\

\textbf{``Inverse Solomonoff Induction''.} Suppose that you observe one bit $x_1,x_2,x_3,\ldots$ after the other, distributed according to algorithmic probability $\mathbf{M}(b|x_1 ^n)$ (this is the formalization of algorithmic idealism's Postulate 2 in the bit model). Suppose that $\mu$ is any computable measure over the bit sequences. Then, with probability at least $2^{-K(\mu)}$, we have
\[
   \lim_{n\to\infty} \left| \mathbf{M}(b|x_1^n) - \mu(b|x_1^n)\right| = 0\qquad (b\in\{0,1\}).
\]
$\strut$\\
That is, for every computable measure $\mu$, there is a positive probability that algorithmic probability (and hence the actual chances of what happens to the agent), conditioned on the previous observations, will converge to $\mu$. Not all measures are equally likely, though. Colloquially, we can think of $K(\mu)$ being small (probability $2^{-K(\mu)}$ being large) as saying that there is a probabilistic computational process that generates $\mu$ which has a short program, i.e.\ is simple in this sense. It can be modelled as running on some machine with a dedicated part (for Turing machines\footnote{Algorithmic idealism does not tell you what the computational model is; all computational models can simulate each other and are hence algorithmically equivalent, up to qualifications such as ``being prefix-free'' which have direct impact on the set of all possible algorithmic probability distributions that these machines generate. But whether you, say, describe an algorithm in terms of a Turing machine or of a cellular automaton is completely up to you. In particular, algorithmic idealism does \textit{not} claim that an agent's external world should resemble any of our mathematical models that we happen to have accidentally introduced and studied in computer science.}, the output tape) that contains the self state $x$ which evolves when the computation unfolds. As every computation, it starts in some simple initial state, with a short program of length $K(\mu)$ that tells the machine what to do. After this, it unfolds probabilistically, and it contains \textit{many other things in addition} to the self state (for Turing machines, it would be other tapes, working memory etc.) such that the resulting evolution of the self state can be understood to arise from this in a ``mechanical'' manner.

But this is a very good description of what we mean by our external world $W$: a larger process into which we seem embedded, and which seems to have once been in a simple distinguished state (the Big Bang) and then evolved probabilistically according to simple laws. We can reason about things in our external world (e.g., a brick falling towards my foot) and use this to construct mechanistic, causal explanations for the evolution of our self state (e.g., the sensation of pain and the observation that something red can be seen when I look down). The fact that our world is formally computable in some sense, and can hence colloquially be understood as a computation, is a well-known and widely discussed thesis that comes in different versions~\cite{Schmidhuber,Lloyd}.

Hence, the inverse Solomonoff induction result can be interpreted as saying that with probability at least $2^{-K(\mu)}$, you will find yourself effectively embedded into the world $W$ pertaining to $\mu$ that we have just described. That is, your first-person probabilities $\mathbf{P}_{\rm 1st}(y|x)=\mathbf{M}(b|x)$ of what actually happens to you (where $x=x_1^n$ and $y=xb$) will be very close to the ``third-person probabilities'' $\mathbf{P}_{\rm 3rd}(y|x)=\mu(b|x)$ of how your representation in world $W$ evolves. In the bit model, this will persist in the limit of $n\to\infty$, that is, ad infinitum. In other formalizations of algorithmic idealism that admit ``forgetting'', we expect this to be the case as long as the self state retains enough memory of previous observations that indicate that world $W$ is a good explanation. But this will have to be demonstrated mathematically, which is the subject of ongoing work.

Consequently, you can pretend that you are currently effectively embedded into some world $W$ (intuitively, until you have a dramatic loss of memory), even if you are fundamentally not. Which world $W$ should you initially have expected to be in, had you been able to wonder? According to algorithmic idealism, the answer has not been predetermined. In the bit model, you could have said that algorithmically simpler worlds\footnote{In more detail, it is not only the world $W$ that has to be simple, but also the specification of ``your place in it'', since both are implied by a program that generates the resulting self state distribution $\mu$.}, i.e.\ ones with shorter program length $K(\mu)$, are more likely, which seems to indicate that you should not be surprised to find yourself in a world that can be described in terms of a small set of mathematically simple physical laws. In particular, the picture that there is ``one actually existing, realized world among all the infinitely many possible worlds'' has to go, which implies a view similar to Lewis' modal realism~\cite{Lewis}.

But if our usual ontological picture of reality is demolished, by what is it replaced? The answer will depend on the specific formalization of algorithmic idealism, so let me only give an example of one possibility. Imagine a formalization of algorithmic idealism where the random walk on the self states is an ergodic Markov chain, then you are on an infinite journey on which you visit every self state infinitely many times. You will sometimes land on a self state $x$ with the property that $\mathbf{P}_{\rm 1st}(y|x)\approx \mu(y|x)$ for a probability (or more general likelihood structure) $\mu$ corresponding to some simple world $W$. This will be the case for a few time steps if and only if $W$ (up to minor modifications) is essentially the unique plausible explanation for $x$ that would be found by universal induction. Hence you will find yourself effectively embedded into world $W$ until some dramatic event happens that erases most of your memory, and you will effectively ``fall out of the world''. After more random walking, you will land on another self state with a similar property, and you will find yourself embedded into another (probably simple) world $W'$ --- and so on, ad infinitum.

\subsection{More than one observer: emergent objective reality$\ldots$}
\label{SubsecReality}

The above argumentation may seem suspiciously solipsistic. I have described how an external world may emerge \textit{for a single agent}, but what about other agents? Is algorithmic idealism committed to \textit{solipsism}, in the sense that it postulates the existence of only a \textit{single} observer? Recall that observers or agents are not fundamental elements of algorithmic idealism, but only momentary snapshots of their structure: the self states, which is similar to Parfit's concept of a person stage. Hence, it does not make sense to ask ``how many agents'' there are in algorithmic idealism, and we have to be more precise about the question that we are actually asking when raising this objection.

Here is a meaningful question to ask. Consider an agent (say, \textit{you}) that currently happens to find itself effectively embedded into a simple computable probabilistic world $W$ --- we already understand how and why this can happen. Now, suppose that you identify something in your external world (``my friend Bob$_{\rm 3rd}$'', where the notation indicates that you refer to him from a third-person perspective) that carries a representation of some self state $x$, evolving over time in a way that is broadly consistent with the specific chosen formalization of Postulate 2 (in the bit model, for example, it would have to contain a bit string of non-decreasing length without any erasure or ``forgetting''). Perhaps $x$ encodes Bob$_{\rm 3rd}$'s brain state, and you would intuitively like to conclude that \textit{there is a first-person perspective associated to it}, some actual ``self'' Bob$_{\rm 1st}$ that is associated to $x$. To clarify what we mean by this, let us compare two probability distributions\footnote{For the sake of the argument, we assume that the ``likelihood'' claims of Postulate 2 are formalized in terms of a probability distribution (or several ones, as it is the case in the bit model). Other choices are certainly possible, and my hope is that the gist of this argument applies more broadly.}:
\begin{compactitem}
\item the probability $\mathbf{P}_{\rm 1st}(y_1,\ldots,y_m|x)$ that $x$ will change into $y_1,y_2,\ldots,y_m$ according to algorithmic idealism's Postulate 2;
\item the probability $\mathbf{P}_{\rm 3rd}(y_1,\ldots,y_m|x)$ that determines how the information content of Bob$_{\rm 3rd}$ evolves according to the probabilistic evolution of world $W$.
\end{compactitem}
These probabilities describe the distributions of future observations $y_1,\ldots,y_m$ of an actual agent in self state $x$ (Bob$_{\rm 1st}$), and of the future recordings in the piece of matter in your world that you call Bob$_{\rm 3rd}$. Intuitively, we assume that the first- and third-person descriptions align --- indeed, the prospect that they would \textit{not} align seems terrifying, because it would indicate that your interaction with the other agent (not just with the piece of matter) would in some sense only be an illusion. In particular, this requires $\mathbf{P}_{\rm 1st}\approx\mathbf{P}_{\rm 3rd}$: \textit{if you have a high chance of seeing your friend Bob$_{\rm 3rd}$ see the sun rise tomorrow, then Bob$_{\rm 1st}$ will indeed have a high chance of seeing the sun rise tomorrow}. This implies that we can causally affect Bob$_{\rm 1st}$ by interacting with Bob$_{\rm 3rd}$ (say, Bob will be happy because you hug him), as long as we can have an effective notion\footnote{Even in classical mechanics, we have an effective notion of intervention, e.g.\ by talking about deciding to set particular initial conditions when running an experiment. Determinism (or lawlike probabilistic behavior) should not be seen as a restriction on the possibility of free will as a higher-level phenomenon, as has been argued by compatibilists such as Daniel Dennett~\cite{Dennett}.} of \textit{intervention} in world $W$. In a physicalist worldview, this is of course a triviality, but in algorithmic idealism it is not.

Indeed, algorithmic idealism predicts that $\mathbf{P}_{\rm 1st}\approx \mathbf{P}_{\rm 3rd}$ under natural conditions that we expect to be satisfied when we encounter ordinary observers in our every-day lives, and in this sense it predicts the emergence of objective reality. This can be seen as follows. Due to Postulate~2, $\mathbf{P}_{\rm 1st}$ is exactly what the method of universal induction $M$ would give you. But similarly as in the Earth example above, if $x$ contains enough data (think about a pretty accurate description of all functionally relevant details in your friend Bob's brain), then a good method of induction is expected to extract an exhaustive description of all relevant physical laws of world $W$ from it, and of all relevant facts of Bob$_{\rm 3rd}$'s environment --- and hence is expected to predict that this self state evolves as if it were embedded into world $W$. In other words, this method of inductive inference should approximately predict $\mathbf{P}_{\rm 3rd}$. Thus, according to Postulate 2, what happens to Bob \textit{from his perspective} (via $\mathbf{P}_{\rm 1st}$) will be in agreement with $\mathbf{P}_{\rm 3rd}$: you share the world $W$ with Bob.

In the bit model, the correctness of this argumentation is indeed a theorem: the evolution of your friend Bob's self state according to world $W$ is $\mathbf{P}_{\rm 3rd}$, which is a computable measure because the world $W$ is computable. Hence, due to some variant of Eq.~(\ref{eqSolInd}), algorithmic probability (or rather its normalized version $\mathbf{P}_{\rm 1st}$) must converge towards it with worldly probability one, in the limit of more and more bits that Bob learns.

Returning to the question of the beginning of this section, this result suggests that algorithmic idealism should not be classified as solipsistic: if Alice points at some piece of matter (or structure) that encodes some self state $x$ in her world, then she can in many cases conclude that ``somebody is really at home'' in this structure in some sense. This is because if $\mathbf{P}_{\rm 1st}\approx \mathbf{P}_{\rm 3rd}$, the probabilistic evolution of this structure in Alice's world (via $\mathbf{P}_{\rm 3rd}$) faithfully represents the evolution of some Bob's first-person perspective (via $\mathbf{P}_{\rm 1st}$), even if Alice interacts with it.

More formally, the question ``how many agents are there?'' (one, many, infinitely many?) is not well-defined in algorithmic idealism, because only self states and their probabilistic transitions are fundamental and well-defined, while the notion of ``agent'' is a pragmatic and vague term that will only sometimes work as expected. More fundamentally, the metaphysical basis of even raising the charge of solipsism is denied to apply in algorithmic idealism: in this view, there are infinitely many self states, they certainly all exist mathematically, but ``physical existence'' is not a meaningful property that a given self state could lack or possess (the trivial observation notwithstanding that a given observer may only see some of these self states in its own emergent external world). While algorithmic idealism hence rejects the basis of the usage of the word ``exist'' in the question ``Do other minds exist?'', the in spirit best answer to this formally undefined question is to say that algorithmic idealism suggests the existence of all possible minds in some sense, and hence is maximally non-solipsistic. 

\subsection{$\ldots$ and its limitations: probabilistic changelings and Boltzmann brains}
The result of the previous subsection hinges on an important assumption: that the other agent Bob's self state is complex enough to make universal induction infer Alice's probabilistic world $W$ from it, so that it is also Bob's world. If this is not the case (in particular, in the bit model, if $K(x)\ll K(\mathbf{P}_{\rm 3rd})$, where $x$ is Bob's current self state, see~\cite{Mueller2020}), then we will in general have $\mathbf{P}_{\rm 1st}(y|x)\not\approx \mathbf{P}_{\rm 3rd}(y|x)$. Then we have an extremely counterintuitive phenomenon: the agent will be a \textit{probabilistic changeling}\footnote{In earlier work such as~\cite{Mueller2020}, probabilistic changelings have been called \textit{probabilistic zombies}, but this seems to convey the wrong types of intuitions, hence the change of name.}. In medieval European superstition, the changeling was an infant that was foisted on a recent mother by a demonic being in exchange for her own child with the intention of harassing and harming people; while the mother \textit{thought} she would have her own child at home, it was \textit{actually} some other being. To see why this is sort of a fitting (albeit limited and morbid) analogy, consider the following example.

As before, suppose that you are currently effectively embedded into some simple probabilistic world $W$ (you probably are as you are reading this), but suppose it is the year 2178 and we have science fiction technology: a computer on your desk simulates another simple probabilistic world $W'$ that has given rise to the evolution of intelligence. For a while, you have been observing one of the simulated beings, which you call Charlie$_{\rm 3rd}$. Given the momentary self state $x$ encoded into Charlie$_{\rm 3rd}$, you know that it is complex enough to make the universal method of induction infer world $W'$ from it\footnote{Note that the method of induction will infer world $W'$ \textit{up to predictive equivalence}, i.e.\ it might also infer world $W$ with the running simulation. The goal of inductive inference is to predict the future, not to explain it.}. Hence, you conclude that \textit{whoever is that self state} (colloquially, you associate the name Charlie$_{\rm 1st}$ to this agent, even though you know that algorithmic idealism has no fundamental notion of agent) \textit{will indeed continue to experience world $W'$}; formally, this means that $\mathbf{P}_{\rm 1st}(y|x)\approx \mathbf{P}_{\rm 3rd}(y|x)$. In other words, Charlie$_{\rm 3rd}$ is an accurate representation of Charlie$_{\rm 1st}$ --- in this sense, Charlie$_{\rm 1st}$ is ``really at home'' in the simulation.

But now you decide to take a malicious action: you intervene in the simulation and arbitrarily replace some of the features in Charlie's environment. Perhaps you place a giant, pink, simulated rabbit in a spacesuit next to Charlie's house, without anyone in the simulation being able to explain where it comes from. This will dramatically alter the future probabilities $\mathbf{P}_{\rm 3rd}$ for Charlie$_{\rm 3rd}$ (implying, for example, high probability for Charlie$_{\rm 3rd}$ displaying strong surprisal soon). However, it cannot alter the probabilities $\mathbf{P}_{\rm 1st}$ which will continue to determine Charlie$_{\rm 1st}$'s future, since these only depend on Charlie's self state $x$ which you have kept intact. In some sense, you have replaced simple world $W'$ on your desk with some other, much more algorithmically complicated world $W''$, but Charlie$_{\rm 1st}$ will probably continue to experience living in world $W'$. While you will continue to observe Charlie$_{\rm 3rd}$ in the simulation, your beloved Charlie$_{\rm 1st}$ is gone --- and has been replaced by an unlikely Changeling$_{\rm 1st}$ who is now realized in Charlie$_{\rm 3rd}$.

You may now feel sad because you really liked Charlie$_{\rm 1st}$ and she is gone, but it would be wrong to think that Changeling$_{\rm 1st}$ is somehow not a likable agent, or an unconscious zombie of some sort. If you thought that Charlie$_{\rm 1st}$ is conscious, then you have no reason to doubt that Changeling$_{\rm 1st}$ is conscious too, and she will very strongly feel that her past was accurately described by Charlie$_{\rm 3rd}$ previously, and hence she will feel like a legitimate successor of Charlie$_{\rm 1st}$ as you knew her before the malicious rabbit event. It is simply that an agent on its random walk on the self states will less often end up in world $W''$ because it is algorithmically more complicated; it corresponds to a path less trodden in computable possibilityspace. We have a counterintuitive situation where chances of events are strongly agent-relative, and where the notion of ``agent'' gets dissolved in ways that are even more enigmatic than, say, in Everettian Quantum Theory. But even if this phenomenon is extremely counterintuitive, it is mathematically consistent.

This curious phenomenon is at the heart of algorithmic idealism's resolution of exotic puzzles such as the Boltzmann brain problem. A mathematically and information-theoretically detailed discussion of this problem within the bit model is given in~\cite{Mueller2020}, and about its relation to Restriction A in~\cite{JonesMueller}; here we only sketch the main idea of a treatment that should apply to all models of algorithmic idealism. Suppose that the world $W$ into which you happen to be effectively embedded is indeed combinatorially large, and you know that there must be many more Boltzmann brains than ordinary ones. Consider some ordinary brain OB and some Boltzmann brain BB that both encode your current self state $x$. Recall that according to algorithmic idealism, it would be a category mistake to say ``I am the OB'' or ``I am the BB'', since you are simply your self state $x$, and this self state is realized or represented both in the OB and in the BB. Regardless of what happens to the OB or the BB, what happens to \textit{you} is characterized by $\mathbf{P}_{\rm 1st}(y|x)$. The third-person perspective tells you that the OB and the BB are going to evolve very differently in the near future according to world $W$, that is, $\mathbf{P}_{\rm 3rd}^{\rm OB}(y|x)\not\approx \mathbf{P}_{\rm 3rd}^{\rm BB}(y|x)$. But then, these probabilities cannot both be approximately equal to $\mathbf{P}_{\rm 1st}(y|x)$, and the ordinary brain or the Boltzmann brain (or both) must become a probabilistic changeling.

In fact, it will be the Boltzmann brain who will become the probabilistic changeling, while $\mathbf{P}_{\rm 1st}(y|x)\approx\mathbf{P}_{\rm 3rd}^{\rm OB}(y|x)$. An explicit technical demonstration of this in the bit model is given in~\cite{Mueller2020}. Intuitively, this is again due to Postulate 2: what happens to you next is what a universal method of induction would predict. Inductive inference tends to extrapolate regularities in your current self state (perhaps contained in your memory of what you interpret as a large number of past observations) into the future, including the pattern of looking as if you had been immersed in a low-entropy planet-like environment with a myriad of regularities. In particular, for the bit model, $\mathbf{P}_{\rm 1st}(y|x)$ is a version of algorithmic probability, and its value being large means essentially that a small amount of algorithmic information is needed to compute $y$ from $x$, regardless of how many instances or copies of $x$ there might be in the external world that is inferred from $x$.

In summary: even if you have reasons to believe that your external world contains many more Boltzmann brains than ordinary brains, what will happen to you next will most likely still be what you would expect if you believed that you are an OB, not a BB. As explained above, algorithmic idealism denies that the question \textit{``Am I an ordinary brain or a Boltzmann brain?''} makes literal sense as it stands: you are your self state, and not one of its realizations. But it predicts that your future experiences will with overwhelming probability confirm ``business as usual on Earth'' rather than Boltzmann brain flamboyancy --- regardless of the actual \textit{number} of Boltzmann brains in the universe.

\section{Example: how to think about the simulation hypothesis}
\label{SecSimHypothesis}

The following five propositions jointly entail that you are probably in a simulation:\\

\begin{compactenum}
\item[1)] It is possible to simulate conscious minds.
\item[2)] Technological progress will not stop any time soon.
\item[3)] Many advanced civilizations do not destroy themselves.
\item[4)] Many advanced civilizations want to run simulations.
\item[5)] If there are many more simulated beings than unsimulated, you are probably simulated.
\end{compactenum}
$\strut$\\
The argument for this entailment claim is well-known~\cite{BostromSimulation} and will not be rehearsed here. It is relevant here because if one is a physicalist, then it is not at all clear which proposition should be rejected to avoid the conclusion. 

The simulation hypothesis is similar to the Boltzmann brain hypothesis. In the latter, a set of plausible physical assumptions leads to the conclusion that most minds are Boltzmann brains; in the former, a set of plausible technological assumptions leads to the conclusion that most minds are simulations. It might seem that the latter conclusion is not so bad to embrace~\cite{Chalmers}. But in fact it raises similar issues to the former~\cite{Schwitzgebel}, since the most common simulations might be ones that are suddenly and unexpectedly shut off.

But it is not clear which of the five propositions one should reject. From a traditional physicalist perspective, Proposition 5) is difficult to deny as it seems based on a fairly innocent principle of indifference. Proposition 4) is very difficult to deny, since even if some of our descendants choose not to simulate conscious minds given the opportunity, surely many still will. Denying propositions 3) or 2) seems awfully pessimistic. That leaves us with Proposition 1). Some do deny it and argue that consciousness is based in biology, and so cannot be realized in silicon~\cite{Block}. However, this is a minority view that conflicts with functionalism in the Philosophy of Mind as well as prominent neuroscientific theories such as Global Workspace Theory and Integrated Information Theory. 

Algorithmic idealism does not change the assessment of the technology-related Propositions 2), 3) and 4) as compared to the traditional physicalist. But what about Proposition 1)? To approach the answer, let us introduce another proposition:\\
\begin{compactenum}
\item[1')] It is possible to give biological birth to conscious minds.
\end{compactenum}
$\strut$\\
Algorithmic idealism claims that
\begin{quote}
Propositions 1) and 1') have identical truth values. In particular, they are both true if the terms in these propositions are understood correctly.
\end{quote}
Traditional physicalists who are also functionalists about consciousness would arguably agree on this. However, those scholars and algorithmic idealists would have \textit{different reasons} for declaring 1) and 1') equivalent. Here is what they could say:
\begin{quote}
\textbf{Physicalist \& functionalist:} ``Biological birth'' or ``starting a simulation'' are events that \textit{give rise} to a conscious being. That is, they are the material causes of this specific instance of consciousness coming into actual existence at this time and place in the universe, and this specific conscious being would have never existed if this specific event (or another, very similar event) had not happened.

Since biological and simulated minds are functionally equivalent, there is no reason to speculate whether 1) might be true but 1') might be not.

\textbf{Algorithmic idealist:} You are your self state, and if you are conscious then this is a property of your self state. Consequently, and more generally, all self states that deserve being called ``conscious'' exist as mathematical structures (as all unconscious ones do), and they experience transitions according to Postulate 2, \textit{regardless of whether they are actualized in our world}. Hence, ``giving birth'' or ``starting a simulation'' do not \textit{create} a self state, but they create material conditions that \textit{represent} it.

If $x$ is not a probabilistic changeling (recall Section~\ref{SecPredictions}), then it is in some sense correct to say that the material conditions so created have introduced an instance of consciousness into our world. But this should not be understood as ``having created'' a conscious being that would not have otherwise existed, but rather as having enabled the conditions to ``share the world'' with it. The consequences of birth or the start of the simulation are structural: all probabilities for what happens to the material conditions that represent $x$ (described by $\mathbf{P}_{\rm 3rd}(y|x)$) are now very close to the probabilities for what happens to $x$ from its private perspective (described by $\mathbf{P}_{\rm 1st}(y|x)$).
\end{quote}
In particular, algorithmic idealists and physicalists who are also structuralists agree on 1) being true, even though they understand the meaning of this proposition differently.

Finally, what about Proposition 5)? We will see that there is, finally, disagreement between the two camps, for similar reasons as in the Boltzmann brain problem. Let us begin by describing why  physicalists may be inclined to think that 5) is true, even though they perhaps should not if their aim is to be carefully consistent overall.

Consider a universe that is simulation-dominated, i.e.\ there exist many more simulated than biological conscious beings. Physicalists would say that this description captures all there is to say about the physical facts, and all meaningful beliefs to be held are beliefs about facts of the world. But then, almost by definition, physics cannot tell you what you \textit{should} believe about being either one or the other --- we have an instance of \textit{Restriction A}, as introduced in Section~\ref{SecRestrictionA}. In some situations in thermodynamics, it turns out that principles of indifference lead to empirically successful predictions, and they seem sort of intuitively plausible (and you can waive your hands and appeal to maximum entropy or symmetry principles or whatever). Physicalists might want to refer to such arguments, or simply to the feeling that maximal ignorance is the best option if you \textit{must} choose to believe something when, actually, you shouldn't. Hence, as the argument goes, if the universe is simulation-dominated, then you should believe that you are in a simulation.

Algorithmic idealism disagrees, and does so along similar lines as in the Boltzmann brain problem. First of all, it denies that the statement \textit{``I am actually contained in a simulation''} is, as a factual claim, meaningful. You are your self state $x$, and you are fundamentally \textit{unembedded}, even though $x$ is represented in many possible computable worlds, and in many simulations running in those worlds.

However, a suitably operationalized version of the statement above \textit{does} make sense in some cases. Namely, which of these representations of your self state $x$ (if any) tracks more accurately your actual first-person probabilities? For example, you can think of two possible scenarios, which you might colloquially want to describe as follows (though algorithmic idealism warns you that you should not unterstand these colloquial descriptions too literally):
\begin{compactenum}
\item You will continue to experience (what you think is) ``base reality'' as usual (more generally, your complete self state, including all its unconscious aspects, will probabilistically evolve as it would in base reality).
\item In a few seconds from now, you make some unusual simulation-indicating experience: perhaps the Moon suddenly disappears, and is replaced by a face of some strange creature that begins to laugh and explains that you are in a simulation, and that it is its actual creator.
\end{compactenum}
Algorithmic idealism has something to say about the relative probabilities of these scenarios: whatever induction predicts is more likely, \textit{regardless of the number of representations that you may count}. Comparing these two very specific scenarios, we would conclude that the first one is more likely. Note that the same argumentation applies not just to a specific choice of surprising event (such as the above), but to the \textit{general} event that something non-base-reality-conforming will happen, which is predicted to be unlikely. Clearly, this intuitive conclusion has to be rigorously proven, given a concrete mathematical model of algorithmic idealism. The calculation in~\cite{Mueller2020} for the Boltzmann brain problem should yield some guidance on how to do this in the bit model.

In summary, \textbf{algorithmic idealism rejects Premise 5):} even if there are many more simulated civilizations than unsimulated ones, this does not imply that you should think that you are probably simulated. It is not the number of the realizations that is relevant, but what a universal method of inductive inference would predict based on your current self state. Note that this is not an \textit{epistemic} statement, i.e.\ merely a description of what some method of induction gives you, but a \textit{lawlike} claim that you will \textit{actually probably} not observe any unusual simulation-indicating experiences in the future.

It is important to note that algorithmic idealism does \textit{not} tell you without further qualification that you are probably not simulated. For instance, it claims that there is no ontological difference between the following two claims:
\begin{compactitem}
    \item[(a)] You will continue to experience base reality because you are actually embedded in base reality.
    \item[(b)] You will continue to experience base reality because you are embedded in a very accurate simulation of it.
\end{compactitem}
According to algorithmic idealism, you are \textit{unembedded}, and hence both statements are equally meaningless: their formulation contains a category mistake. In particular, there is no sense in which any one of them would be ``more correct'' than the other.

Recall that algorithmic idealism predicts that agents will often find themselves \textit{effectively embedded} into some computable world $W$: pretending that you are embedded will accurately track your first-person chances. It is conceivable that your self state is represented in $W$ within something that you might call a computer simulation running in $W$. However, if this simulation (call it $W'$) is \textit{isolated} --- that is, if it has been programmed and keeps running without any intervention from the outside --- then it is equally correct to pretend that you are embedded into standalone world $W'$ (without the world $W$ and computer running it). Hence, the only meaningful instance in which algorithmic idealism gives you an effective embedding \textit{specifically into a simulation} is in cases where the simulation is not isolated, but information from the external world $W$ leaks into the simulation (for example by intervention). This distinction, together with further implications for the simulation of agents (including ethical questions raised by Bostrom~\cite{Bostrom}) is discussed in more detail in~\cite{Mueller2020}.

As a final remark, let us get back to a comment of the beginning of this section that articulated a particular worry against the simulation hypothesis: most common simulations might be suddenly and unexpectedly shut off. Does this worry not indicate a fundamental difference between the two scenarios (a) and (b) above? Algorithmic idealism denies this. Recall the ``changeling'' example of Section~\ref{SecPredictions}, where you run an isolated simulation of a world $W'$ which contains an agent Charlie$_{\rm 3rd}$. Suppose that you decide to shut the simulation down. This means that Charlie$_{\rm 3rd}$ (the corresponding pixels shown on your screen, or the pattern in the computer memory, if you wish) ceases to exist. However, Charlie$_{\rm 1st}$'s probabilities $\mathbf{P}_{\rm 1st}(y|x)$ of her future observations $y$ are essentially\footnote{In the case of the bit model, it should be considered to be modified by an unimaginably tiny amount: among all supporting explanations for predicting $y$ according to $W'$-probability, the contribution coming from ``world $W$ simulating $W'$'' disappears, and hence universal induction may be ever so slightly less inclined to converge to this particular probability value.} unaltered by this: the universal method of induction (mentioned in Postulate 2) will continue to infer that world $W'$ is essentially the best explanation for Charlie$_{\rm 1st}$'s self state, and hence Charlie$_{\rm 1st}$ will continue to experience world $W'$. According to algorithmic idealism, shutting down a closed simulation does not ``kill'' its inhabitants, but it leaves them essentially unaffected. Intuitively, your simulation is more of a ``movie'' than a ``zoo'', and hence its destruction does not affect its protagonists. For science fiction fans, this has amusing parallels with the ``dust theory'' described in Greg Egan's \textit{Permutation City}~\cite{PermutationCity}.

\section{Conclusions}
I have presented an approach to the Foundations of Physics, tentatively called algorithmic idealism, which declares the first-person perspective fundamental. Essentially, it is the question of ``what will probably happen to me next?'' that is taken to be fundamental, rather than the question of ``what is the case in the world?''. Despite its name, high-level and vague concepts such as agents, consciousness or belief do not play any fundamental role in algorithmic idealism, and its two postulates admit several possible rigorous mathematical formalizations, i.e.\ a class of concrete theories. One preliminary formalization, the bit model, has been presented in earlier work~\cite{Mueller2020}, which also contained formal proofs of many of the conceptual arguments that I have put forward here about what to expect from an approach like this: the emergence (not exactly, but to good approximation!) of a simple, probabilistic, computable external world into which the agent seems to be embedded, and of a notion of objective reality; the dissolution of enigmas such as the Boltzmann brain problem, Parfit's teletransportation paradox, or the computer simulation of agents; and novel phenomena such as probabilistic changelings, implying a particular agent-dependency\footnote{I wholeheartedly and passionately oppose attempts to read this as supporting esoteric nonsense. In every-day life and almost all situations in physics, facts are observer-independent. Claims are true or false, regardless of who expresses them~\cite{SokalBricmont}, and truth exists and is observer-independent. The sort of agent-dependency that I am describing here is more of the sort of observer-dependency of lengths and time durations in special relativity: hard mathematical relations, rigorously formalized, and independent of what we wish, feel or believe.} of facts and their likelihood.

By construction, the predictions of algorithmic idealism agree with those of our physical theories in standard laboratory situations if a suitable physical version of the Church-Turing thesis is true. Indeed, algorithmic idealism is constructed in the spirit of empiricism: no insights of any kind into the ontology of the world are directly sought, and essentially all human preconceptions about metaphysics are demolished on the way (or, rather, declared to be only useful models that sometimes give approximately the right answers). Rather than asking what the world is like, the sole goal of this approach is to get the phenomena right. In this sense, it intends to represent a minimal extension of the usual methodology of physics into new territory: from the regime of \textit{intersubjective experiments}, where everybody can learn about the results, to \textit{private experiments} for which a third-person view or prediction ``from the outside'' is in principle unavailable.

The mere \textit{possibility} (even if considered unlikely) that some aspects of algorithmic idealism are closer to the truth than our standard view hints at a surprising lack of imagination in the way we look at some essential questions of humanity. Should humanity colonize the galaxy in order to make a large number of conscious lives possible that would not otherwise have existed~\cite{Ord}? Arguments in favor of this typically implicitly assume a particular metaphysics of ``one single real world among infinitely many possible ones'' that has already been criticized by Lewis~\cite{Lewis} and that is rejected by algorithmic idealism. Does shutting down a computer simulation terminate its simulated conscious beings~\cite{Schwitzgebel}? This usually seems to be taken as obviously true, but algorithmic idealism claims that this is not generally the case, and that the answer will depend on how much information from the outside world is fed into the simulation. But if we do not understand what to say about the simulation of agents, how can we be so sure we understand death (recall Section~\ref{SecPrologue})? The mere reason why we consider the view of transitioning into an infinite-duration unconscious ``muted state'' as the only scientifically plausible possibility may be that we have so far only seen alternatives promoting esoteric nonsense and religious wishful thinking, and that we want to signal how serious and trustworthy we are by not even addressing the question.

These examples also demonstrate that algorithmic idealism is not merely a theory of how agents can best predict their future observations, but a radically new approach to physical reality. If it were simply some sort of theory of epistemology, then \textit{what will actually happen} to an agent would be determined by the (in general unknown) physical world in which the agent would operate. In contrast, algorithmic idealism denies that the agent is fundamentally embedded into some world, and grounds whatever will happen to the agent on its self state alone. This counterintuitive starting point allows it to formulate a law (Postulate 2) which is meant to express how an agent's fate is determined in mundane but also truly exotic situations, such as simulation and multiplication scenarios, and it makes very counterintuitive predictions (such as the possibility that agents may ``fall out'' of the world into which they are effectively approximately embedded, or the notion of probabilistic changelings). These predictions would not only be impossible to \textit{obtain} in an epistemic approach (because, as we discussed via Restriction A, third-person claims about the world have nothing to say about first-person chances in, say, duplication scenarios), but they \textit{outright contradict} the assumption of an agent that is embedded into some given but unknown world. This also indicates a clear separation from \textit{operational theories}~\cite{Coecke} as they are often formulated in the quantum information theory context: while these theories also often feature agents and their observations (such as Alice and Bob in a communication scenario), these agents are assumed to be embedded into a joint physical system (external world) that mediates their communication and observations. In algorithmic idealism, on the other hand, this external world is not postulated a priori, but an emergent statistical phenomenon, and so is the intuitive idea that Alice and Bob share the \textit{same} physical world.

In summary, algorithmic idealism was not constructed with the intention to say anything directly about reality. However, the methodological starting point to ignore the usual realist assumptions led to a formulation which indeed admitted novel predictions that are not otherwise available, but at the same time made it incompatible with the received view of physical reality, and thus distinct from existing epistemic or operational approaches.

There are two main open technical problems for algorithmic idealism, and these turn out to be related. First, the bit model of~\cite{Mueller2020} has to be replaced by a better mathematical formalization, and this is currently work in progress. Essentially, the bit model precludes the possibility of ``forgetting'', which is unmotivated, unrealistic, and an artefact of the way that Solomonoff induction has been formulated in Algorithmic Information Theory.

Second, a successful formalization of algorithmic idealism should predict some phenomenal aspects of Quantum Theory. Clearly, Quantum Theory is one of the main motivations for algorithmic idealism, as explained in Section~\ref{SecQuantum}. Moreover, if its probabilities of events are fundamentally agent-relative, then this already resembles a phenomenon that is sometimes considered genuinely quantum (as in the extended Wigner's friend scenario~\cite{Bong} of Section~\ref{SecRestrictionA}). It also hints at the possibility that the resulting description of an agent's external world may have the feature that not all potential events fit into a joint Kolmogorovian probability distribution, a fundamental structural element of Quantum Theory~\cite{Fine} which is necessary (though not sufficient) for many of its phenomena. The main goal should be to \textit{reconstruct}~\cite{Mueller2021,Hardy2001,DakicBrukner2011,MasanesMueller2011,Chiribella2011} phenomena that are genuinely quantum, such as the violation of Bell inequalities together with compliance with the no-signalling principle~\cite{Barrett2005}, or the possibility of performing BQP-complete computations in polynomial time~\cite{Aaronson}. Indeed, a potential explanation for the former phenomenon in models of algorithmic idealism that admit forgetting has been sketched in~\cite{Mueller2020}, but it needs further elaboration. Note that it is \textit{not} enough to point at mathematical elements (such as complex numbers, ``wave functions'' or evolution equations) that resemble mathematical elements of Quantum Theory, as is done in some approaches. Instead, the main question is whether the \textit{statistical results of experiments} are predicted to behave in genuinely nonclassical ways, and any claims of saying something about this must engage with the insights from research on the foundations of Quantum Theory. In particular, its ability (or inability) to predict aspects of Quantum Theory is an important opportunity to test algorithmic idealism, and this alone suggests that it would be methodologically wrong to build the quantum formalism into algorithmic idealism by construction.

There are several other elements of algorithmic idealism that need further elaboration. As a simple example, its reference to a ``physical Church-Turing thesis'' has to be further fleshed out: what exactly is assumed here about the computability of our physical world? Note that only something \textit{much} weaker than Zuse's thesis~\cite{Zuse} and even weaker than Deutsch's version~\cite{Deutsch} is needed such that algorithmic idealism is guaranteed to be consistent with physics: the method of induction $M$ that is used in the formalization of Postulate 2 should be successful in our physical world. In the bit model, $M$ is Solomonoff induction. In a nutshell, for it to be successful, we need the existence of an algorithm which, given a description of an experiment with discrete outcomes, computes the probabilities of all possible results. This seems to be a rather harmless presumption which neither assumes that our physical world is discrete nor that it is deterministic. However, the details of this assumption have to be worked out, and the claim that this applies to our current theories has to be verified in detail, while keeping in mind that the method of induction will change under a modification of the formalization of Postulate 2.

As an epilogue, let us finally return to the scenario of Section~\ref{SecPrologue}: what should we answer the patient's desperate question, under the unlikely assumption that algorithmic idealism is true? Will it ``wake up'' in the computer simulation or not? Given that we do not know the patient's self state, the details of the simulation, and that we do not even rely on a particular formalization of Postulates 1 and 2, we can only attempt some sort of ``rule of thumb'' answer based on general principles of algorithmic idealism: you are your self state, and there is no need to be realized in a material object in what you currently experience as your external world in order to be ``awake''. So yes, perhaps you will experience the simulated environment. But then, the simulation should be \textit{algorithmically simple}, or otherwise the transition will be unlikely --- a complicated paradise-like environment tailored to your needs will probably not do the job. So, in the end, maybe you would not even have needed to spend your money, because the simulated world exists mathematically regardless of what AfterMath has been doing. At the very least, you do not have to pay AfterMath to prevent them from shutting it down later. But to say more, one would need a detailed mathematical model of algorithmic idealism and a lot more work.

The rule of thumb answer can hence perhaps be summarized in two words: \textit{be hopeful}.

\section*{Acknowledgments}
I am grateful to Kelvin McQueen for many helpful and stimulating discussions, and in particular for his earlier contributions to what has become Section~\ref{SecSimHypothesis} of the present paper. I would also like to thank Michael Cuffaro for helpful comments on an earlier draft.

\end{document}